\documentclass{aa}
\usepackage{graphicx}
\usepackage{natbib}
\begin{document}
\title{The Nitrogen and Oxygen abundances in the neutral gas at high redshift\thanks{Based on 
observations carried out at the European
Southern Observatory (ESO), under visitor mode progs. ID 65.O-0063, 66.A-0624,
67.A-0078 and 68.A-0600 with the UVES echelle
spectrograph installed at the ESO Very Large Telescope (VLT), unit
Kueyen, on mount Paranal in Chile. Also based on archival data from progs. 68.A-0492 
(PI: D'Odorico), 68.B-0115 (PI: Molaro) and 69.A-0051 (PI: Pettini).}
}
   \titlerunning{Nitrogen and Oxygen Abundance Ratio in DLA Systems}

   \author{
          Patrick Petitjean\inst{1}
          \and
          C\'edric Ledoux\inst{2}
        \and
          R. Srianand\inst{3}
           }

   \offprints{Patrick Petitjean, petitjean@iap.fr}

   \institute{Institut d'Astrophysique de Paris, CNRS and UPMC Paris 6, UMR7095,
              98bis Boulevard Arago, F-75014, Paris, France
%         \and 
%              LERMA, Observatoire de Paris, 61 Avenue de l'Observatoire, 
%              F-75014, Paris, France
         \and
             European Southern Observatory, Alonso de C\'ordova 
             3107, Casilla 19001, Vitacura, Santiago, Chile
         \and
             IUCAA, Post Bag 4, Ganesh Khind, Pune 411 007, India 
             }
   
   \date{ }

   \abstract
{}
{We study the Oxygen and Nitrogen abundances in the interstellar
medium of high-redshift galaxies.}
{We use high resolution and high signal-to-noise ratio spectra of
Damped Lyman-$\alpha$ (DLA) systems detected along the line-of-sight to quasars 
to derive robust abundance measurements from unsaturated metal absorption lines.} 
{We present results for a sample of 16 high-redshift DLAs and strong sub-DLAs 
(log~$N$(H~{\sc i})~$>$~19.5, 2.4~$<$~$z_{\rm abs}$~$<$3.6) including
13 new measurements. 
We find that the Oxygen to Iron abundance ratio is pretty much constant with 
[O/Fe]~$\sim$~+0.32$\pm$0.10 for $-2.5$~$<$~[O/H]~$<$~$-1.0$ with a small scatter around 
this value. The Oxygen abundance follows quite well the Silicon abundance within $\sim$0.2~dex
although the Silicon abundance could be slightly smaller for [O/H]~$<$~$-2$.
The distribution of the [N/O] abundance ratio, measured from components that
are detected in both species, 
is somehow double peaked: five systems have [N/O]~$>$~$-$1 and nine systems have [N/O]~$<$~$-$1.15.
%The diagram [N/O] versus [O/H] shows that there is no strong evidence for a plateau 
%at [N/O]~$\sim$~$-$1.4. 
%six systems have [N/O] consistent within errors with values between $-$1.3 and $-$1.4.
In the diagram [N/O] versus [O/H], a loose plateau is possibly present at 
[N/O]~$\sim$~$-$0.9 that is below the so-called primary plateau as seen in local 
metal-poor dwarf galaxies
([N/O] in the range $-$0.57 to $-$0.74). 
%The discrepancy could be due to the overabundance of Oxygen compared to Iron. 
No system is seen above this primary plateau whereas the majority of 
the systems lie well below with a large scatter.
All this suggests a picture in which DLAs undergo successive star-bursts.
During such an episode, the [N/O] ratio decreases sharply because
of the rapid release of Oxygen by massive stars whereas inbetween
two bursts, Nitrogen is released by low and intermediate-mass stars 
with a delay and the [N/O] ratio increases.
}
{}

   \authorrunning{P. Petitjean et al.}
   \maketitle
%
%________________________________________________________________

\section{Introduction}

The production of Nitrogen in stars is the subject of strong interest as 
it is difficult to explain consistently the Nitrogen abundances measured
in different astrophysical environments.
%it is still 
%unclear which stars are involved and how and when, exactly, the transformation
%of Carbon into Nitrogen occurs. 
New abundance measurements in the interstellar medium (ISM) of our Galaxy as well 
as in external galaxies are of prime importance in these discussions.
This could help decide what the sites of the Nitrogen production are. 
%{\bf If it is generally believed that the main contributors are the long-lived
%stars progenitors of Asymptotic-Giant-Branch stars, the contribution
%by massive stars is still uncertain.}
It is generally believed that the main contributors are the
long-lived intermediate mass stars that are progenitors
of Asymptotic-Giant-Branch (AGB) stars. The contribution
from massive stars is still uncertain.
%
%Are the main contributors massive and short-lived stars or the long-lived
%stars progenitors of Asymptotic-Giant-Branch stars ? 
\par
The main nucleosynthetic pathway for the production of Nitrogen 
is the CNO cycle which takes place in the stellar H-burning layers. 
Nitrogen is thought to have both primary and secondary origins depending
on whether the seed Carbon and Oxygen nuclei are produced by the star itself
(primary) or are already present in the interstellar medium from which the star
forms in which case Carbon and Oxygen seeds are left-overs from previous generations of stars
(secondary). Secondary production can happen in the H-burning layers of all stars
as the Carbon seed is present, by definition, in these layers.
Primary Nitrogen is produced when Carbon, synthetized in the Helium-burning 
shell of the star, penetrates into the Hydrogen-burning upper shell where 
it is transformed in Nitrogen by the CNO cycle. This happens 
in intermediate mass stars (4~$\leq$~$M/M_{\odot}$~$\leq$~7) during
the Asymptotic Giant Branch phase (Henry et al. 2000).
Large uncertainties affect the theoretical predictions of both the primary and
secondary nucleosynthesis of Nitrogen in low and intermediate-mass stars and the possible
contribution of primary Nitrogen from massive stars (Woosley \& Weaver 1995, Marigo 2001,
Maeder \& Meynet 2002).
\par
It is believed that the determination of the Nitrogen abundance in objects with 
metallicities spread over a large range may help answering this question.
Therefore, it is usual to derive metallicities in nearby
low-metallicity emission line galaxies (e.g. Nava et al. 2006;
Izotov et al. 2006) or low-metallicity stars in our Galaxy (Spite et al. 2005).
A complementary approach is to derive metallicities directly at high redshift in
Damped Lyman-$\alpha$ (DLA) systems (Pettini et al. 2002, Prochaska et al. 2002, 
Centuri\'on et al. 2003) that have metallicities typically in the range
$-$2.5~$\leq$~$Z/Z_{\odot}$~$\leq$~$-$1. Due to the large \ion{H}{i} column densities
(log $N$(H~{\sc i})~$\geq$~~10$^{20}$~cm$^{-2}$) and the conspicuous presence of metals, DLAs
are believed to arise in high-redshift galaxies or at least to be located close
to regions where star-formation occurs. 
\par
It is usual to discuss these issues using the diagram giving the Nitrogen to Oxygen 
abundance ratio versus the Oxygen abundance.
In the case of secondary production, the ratio of
Nitrogen to Oxygen abundances increases with increasing Oxygen metallicity; 
whereas for primary production, the ratio remains contant as Nitrogen tracks
Oxygen. In H~{\sc ii} regions of nearby galaxies a trend in the [N/O] abundance ratio is 
seen with a primary plateau at low Oxygen abundances and a secondary behavior for
abundances [O/H]~$>$~$-1$ (van Zee et al. 1998; Izotov \& Thuan 1999). 
%It is interesting to compare in this diagram the points from the local universe
%with those obtained at high redshift in DLA systems.
DLA systems have a very different behavior in this diagram with 
[N/$\alpha$] (see below) measurements well below the primary plateau (Pettini et al. 2002;
Centuri\'on et al. 2003). This is probably related to the star-formation history
of these objects.
% observed in QSO spectra. that are characterized by neutral hydrogen column densities
%larger than log $N$(H~{\sc i})~$\sim$~~10$^{20}$~cm$^{-2}$. 
\par
Oxygen and Nitrogen abundances can be derived directly from the 
$N$(O~{\sc i})/$N$(H~{\sc i}) and $N$(N~{\sc i})/$N$(H~{\sc i}) column density ratios.
Because of efficient charge exchange reactions that link the neutral species together, 
the ionization corrections are negligible for log~$N$(H~{\sc i})~$>$~19.5 (Viegas 1995).
Accurate O~{\sc i} and N~{\sc i} column densities are difficult to derive 
however. The main reason is that the absorption lines are located in the
Lyman-$\alpha$ forest and are often blended. In addition, and this is especially
true for O~{\sc i}, absorption lines are easily saturated. For N~{\sc i} the situation
is less dramatic because the Nitrogen abundance is smaller and N~{\sc i} has
two triplets around $\lambda$1134 and $\lambda$1200~\AA. On the contrary, the main O~{\sc i} 
absorption feature at $\lambda$1302 is a single line and is often saturated.
Other weaker absorption lines are found much further in the blue, therefore deep in the
Lyman-$\alpha$ forest and in a wavelength range where good signal-to-noise ratio (SNR) is difficult
to obtain. This is why, in previous studies, the Oxygen abundance has often been
replaced by that of another $\alpha$-element such as Sulphur or Silicon (Pettini et al. 2002, 
Centuri\'on et al. 2003, Henry \& Prochaska 2007). 
Using Silicon instead of Oxygen, Centuri\'on et al. (2003) 
claimed that 75\% of DLA systems show a mean value [N/Si]~=~$-$0.87 with 
0.17 dex dispersion, corresponding approximately to the primary plateau observed
locally, and 25\% are clustered around [N/Si]~=~$-$1.5 with even less 
dispersion (0.05 dex), the transition between low and high [N/Si] values happening 
at [N/H]~=~$-$2.8 (note that the latter transition may occur at [N/H]~=~$-$3, see
Molaro et al. 2003).
\par
Here we present the (N/O) vs (O/H) diagram using robust determinations
of Oxygen and Nitrogen metallicities derived from unsaturated absorption lines 
selected from the largest UVES sample of DLA systems 
available up to now (Ledoux et al. 2003, 2006a). 
In Section~2 we present our sample. 
We describe the O and N abundance measurements in Section~3 and 
discuss the results and implications in Sections~4 and 5.

%There is a general consensus that Oxygen is almost entirely produced
%by massive Type II SNe 
%primarily during the central hydrogen burning.

%----------------------------------------------------------------

%The data have been reduced using the latest version of the UVES pipeline
%(Ballester et al. 2000) which is available as a dedicated context of the ESO
%MIDAS data reduction system. The main characteristics of the pipeline are to
%perform a precise inter-order background subtraction for science frames
%and master flat-fields, and an optimal extraction with gaussian modelling of
%the object spatial profile rejecting cosmic ray impacts and subtracting the sky
%spectrum simultaneously. The pipeline products were checked step by step.
%The wavelength scale of the spectra reduced by the pipeline was then
%converted to vacuum-heliocentric values and individual 1-D spectra were scaled,
%weighted and combined altogether to produce the final science spectrum and its
%associated variance.
  
   \begin{table*}
      \caption[]{Results of Voigt-profile fitting}
         \label{table1}
          \centering
          \begin{tabular}{l c r r r r r r r r} 
          \hline\hline
Name & n &  $z_{\rm abs}$ & log~$N$(H~{\sc i}) & log~$N$(O~{\sc i}) & $\sigma$ & [O/H] & log~$N$(N~{\sc i}) 
& $\sigma$ & [N/H] \\
     &   &                &  (cm$^{-2}$)         & (cm$^{-2}$)          &          &       &  (cm$^{-2}$) 
&          &       \\
\hline  
%Q 0000$-$263$^a$&1   & 3.39013 & 21.40$\pm$0.08 & 16.42 & 0.10 & $-$1.67 &  14.73 & 0.02 & $-$2.50 \\
Q 0102$-$190    &1   & 2.92646 & 20.00$\pm$0.10 & 14.68 & 0.16 &         &  12.67 & 0.11 &         \\
                &2   & 2.92661 &                & 14.66 & 0.17 &         &$<$12.35& $^d$ &         \\
                &3   & 2.92727 &                & 14.20 & 0.02 &         &  12.38 & 0.09 &         \\
                &4   & 2.92771 &                & 14.41 & 0.02 &         &  12.55 & 0.06 &         \\
                &Tot &         &                & 15.13 & 0.08 & $-$1.56 &  12.76 & 0.11 & $-$3.07 \\
Q 0112$-$306   &1   & 2.41844 & 20.50$\pm$0.08 & 14.78 & 0.12 &         &  12.66 & 0.08 &         \\
                &2   & 2.41861 &                & 14.45 & 0.06 &         &  12.99 & 0.04 &         \\
                &Tot &         &                & 14.95 & 0.08 & $-$2.24 &  13.16 & 0.04 & $-$3.17 \\
Q 0347$-$383$^a$&1   & 3.02463 & 20.73$\pm$0.05 & 16.12 & 0.12 &         &  14.15 & 0.03 &         \\
	    	&2   & 3.02485 &                & 16.18 & 0.18 &         &  14.47 & 0.03 &         \\
                &Tot &         &                & 16.45 & 0.11  & $-$0.97 &  14.64 &0.02 & $-$1.92 \\
Q 0841+129      &1   & 2.47622 & 20.80$\pm$0.10 & 16.09 & 0.03 & $-$1.40 &  13.94 & 0.02 & $-$2.69 \\
Q 0913+072$^b$  &1   & 2.61829 & 20.35$\pm$0.10 & 14.08 & 0.02 &         &$<$12.10& $^d$ &         \\ 
                &2   & 2.61844 &                & 14.43 & 0.02 &         &$<$12.20&      &         \\
                &Tot &         &                & 14.59 & 0.02 & $-$2.45 &$<$12.45&      &$<-$3.73 \\
Q 1108$-$077    &1   & 3.60767 & 20.37$\pm$0.07 & 15.37 & 0.03 & $-$1.69 &$<$12.84& $^d$ &$<-$3.36 \\
Q 1337+113      &1   & 2.50793 & 20.12$\pm$0.05 & 14.90 & 0.10 & $-$1.91 &  12.84 & 0.10 & $-$3.11 \\
Q 1337+113      &1   & 2.79557 & 21.00$\pm$0.08 & 15.21 & 0.12 &         &$<$12.40& $^d$ &         \\
                &2   & 2.79581 &                & 15.59 & 0.08 &         &  13.98 & 0.03 &         \\
                &Tot &         &                & 15.74 & 0.07 & $-$1.95 &  13.99 & 0.03 & $-$2.84 \\
Q 1340$-$136    &1   & 3.11835 & 20.05$\pm$0.08 & 15.52 & 0.02 & $-$1.22 &  13.28 & 0.02 & $-$2.60 \\
Q 1409+095$^c$  &1   & 2.45595 & 20.53$\pm$0.08 & 15.05 & 0.02 &         & $<$13.26&$^d$ &         \\
                &2   & 2.45644 &                & 15.00 & 0.02 &         & $<$13.14&     &         \\ 
                &Tot &         &                & 15.33 & 0.02 & $-$1.89 & $<$13.51&     &$<-$2.85 \\ 
Q 1409+095$^c$  &1   & 2.66821 & 19.80$\pm$0.08 & 15.18 & 0.02 & $-$1.31 &  13.50 & 0.02 & $-$2.13 \\
%Q 1946+76$^e$  &    & 2.84400 & 20.27$\pm$0.06 &       &      & $-$2.19 &        &      & $-$3.66 \\ 
Q 2059$-$360    &1   & 3.08262 & 20.98$\pm$0.08 & 15.63 & 0.06 &         &  13.49 & 0.02 &         \\
                &2   & 3.08293 &                & 15.85 & 0.05 &         &  13.71 & 0.02 &         \\
                &3   & 3.08316 &                & 15.02 & 0.12 &         &  12.85 & 0.14 &         \\
                &Tot &         &                & 16.09 & 0.04 & $-$1.58 &  13.95 & 0.02 & $-$2.86 \\
Q 2332$-$094    &1   & 3.05633 & 20.50$\pm$0.07 & 14.97 & 0.02 &         &$<$12.93& $^d$ &         \\
                &2   & 3.05658 &                & 14.46 & 0.03 &         &$<$12.93& $^d$ &         \\
                &3   & 3.05677 &                & 14.71 & 0.07 &         &$<$12.93& $^d$ &         \\
                &4   & 3.05690 &                & 15.32 & 0.02 &         &$<$12.93& $^d$ &         \\
                &5   & 3.05723 &                & 15.67 & 0.04 &         &  13.73 & 0.03 &         \\
                &6   & 3.05737 &                & 14.66 & 0.04 &         &$<$12.93& $^d$ &         \\
                &Tot &         &                & 15.95 & 0.02 & $-$1.24 &  13.73 & 0.03 & $-$2.60 \\
%QXO 0001$^e$    & & 3.00000 & 20.70$\pm$0.06 &       &      & $-$1.81 &        &      & $-$3.26 &      &        \\ 
         \hline
\multicolumn{10}{l}{$^a$also Ledoux et al. (2003); $^b$ also Erni et al. (2006);  
$^c$ also Pettini et al. (2002); $^d$ $3\sigma$ detection limit.}
         \end{tabular} 
%      \begin{list}{}{}
%%     \item[$^{\mathrm{a}}$] also Molaro et al. (2000)
%     \item[$^{\mathrm{b}}$] also Ledoux et al. (2003)
%\item[$^{\rm c}$] also Erni et al. (2006)
%\item[$^{\rm d}$] $3\sigma$ detection limit.
%\item[$^{\rm f}$] also Pettini et al. (2002)
%     \end{list}
  \end{table*}

\section{UVES DLA sample}\label{sample}
Most of the systems in our sample were selected from the follow-up of the Large
Bright QSO Survey (Wolfe et al. 1995) and observed at the VLT
with UVES between 2000 and 2004 in the course of a systematic search
for molecular hydrogen at $z_{\rm abs}>1.7$ (Petitjean et al. 2000; Ledoux et al. 2003). 
Our sample comprises
61 bona-fide DLA systems ($\log N($H\,{\sc i}$)\ge 20.3$) and 13
strong sub-DLA systems with total neutral hydrogen column densities
in the range $19.5\la \log N($H\,{\sc i}$)<20.3$. Characteristics of the systems
(metallicities, $N$(H~{\sc i}) column densities, kinematics), except
for four sub-DLAs with log~$N$(H~{\sc i})~=~19.7-19.8), are given in
Ledoux et al. (2006a). The absorption line analysis was performed in an homogeneous manner
using standard Voigt-profile fitting techniques adopting the
oscillator strengths compiled by Morton (2003). Total column
densities were derived as the sum of the column densities measured in
individual components of the line profiles. Average gas phase metallicities
relative to solar, 
[X/H$]\equiv\log [N($X$)/N($H$)]-\log [N($X$)/N($H$)]_\odot$,
were calculated using solar abundances from Lodders (2003).
%A noticeable property of this large dataset is that it samples well both 
%ends of the DLA metallicity distribution, from [X/H$]\approx -2.6$ up to 
%about half of Solar.

As we are interested in measuring the [O/H] and [N/H] abundances from the 
$N$(O~{\sc i}), $N$(N~{\sc i}) and $N$(H~{\sc i}) column densities,
we have to select systems where we can derive accurate column densities, e.g.
where at least one transition from both N~{\sc i} and O~{\sc i} is not strongly saturated.
Note that the transitions from these elements are most of the time redshifted in the
Lyman-$\alpha$ forest and we require in addition 
that at least one unsaturated transition be free of blending. 
%We must be confident that the ionization conditions in the gas are such
%that the column density ratios are a good indicator of the abundance ratios, e.g.
%although O~{\sc i} and N~{\sc i} are both tied to H~{\sc i} by charge exchange 
%reactions, this condition is fullfilled only if log $N$(H~{\sc i})~$>$~19.5
%(Viegas 1995). 
From the above sample and under these conditions, 
we selected 13 sytems where we could measure $N$(O~{\sc i}) and
$N$(N~{\sc i}) column densities accurately. 

From the literature, we added three measurements at $z_{\rm abs}$~=~3.390, 2.844 and 
3 toward, respectively, Q~0000$-$263 (Molaro et al. 2001), Q~1946+769 and QXO~0001 
(Prochaska et al. 2002). 
As a consequence of our selection criteria, we did not include the system at 
$z_{\rm abs}$~=~4.466 toward Q~0307$-$4945 (Dessauges-Zavadsky 
et al. 2001) as the complex O~{\sc i} and N~{\sc i}
velocity profiles, spread over more than 200 km~s$^{-1}$,
are highly saturated and/or blended.
Neither did we include the systems at $z_{\rm abs}$~=~2.076 and 2.456 toward, respectively,
Q~2206$-$199 and Q~1409+095 (Pettini et al. 2002) because the Oxygen abundance can only
be ill-defined from strongly saturated O~{\sc i}$\lambda$1302 absorption lines.
In the case of Q~2206$-$199, the strongest N~{\sc i} feature is also blended
(Molaro et al. 2003).

Ionization may be a concern when deriving the (N/O) abundance ratio from
the $N$(N~{\sc i})/$N$(O~{\sc i}) column density ratio (Viegas 1995, Prochaska et al. 2002).
%This correction should be most of the time less than +0.2 dex. 
If the gas is neutral then charge exchange reactions are fast enough so that 
O~{\sc i} and N~{\sc i} are both tied to H~{\sc i} and there is no need for correction for
log~$N$(H~{\sc i})~$>$~19.5 (Viegas 1995). 
%If the gas is not neutral, 
In case of the presence of enough hard photons (or cosmic rays) the ionization balance
could be displaced towards higher ionization species and both O~{\sc i} and N~{\sc i} 
could be over-ionized compared to H~{\sc i}. 
However for most DLA systems, hard photons come predominantly from the background ionization
field. The ionization parameter for this field at $z = 3$ is $U<10^{-4}$ 
for typical densities expected for DLAs ($n > 1$~cm$^{-3}$; see e.g. Petitjean et al. 1992),
and ionization corrections are again small.
% 
%In addition, the ionization correction is very uncertain. The most important reason
%is that the gas along the line of sight is mixed up and the decomposition of
%the absorption blends is not unique. One component can hide an other one and
%it is difficult to assess what is the ionization locally. This is why we have added
%all components together in the system. 
%
Indeed, Prochaska et al. (2002) 
inspected about twenty DLAs and concluded that the ionization correction for [N/$\alpha$] 
is at most of the order of +0.1~dex. This amount should probably be added to the measurement 
errors. In the following we did not apply any correction as we show that
there is no correlation between [N/O] and log~$N$(H~{\sc i}).

\section{Column Densities and Abundances}

We performed in a usual manner Voigt profile fitting of absorption lines associated 
to the 13 systems we selected from our DLA sample (see e.g. Ledoux et al. 2006a, 
Erni et al. 2006).
The overall decomposition in subcomponents is derived from the simultaneous fit
of numerous absorption lines from different species 
(most importantly Zn~{\sc ii}, Fe~{\sc ii} and Si~{\sc ii}) redshifted
outside the Lyman-$\alpha$ forest and free from blending. 
We used the resulting component structure to detect N~{\sc i} and O~{\sc i}
features and confirm that they are not blended. 
Column densities obtained for each of the components are given in Table~1. 
%Results of multi-component Voigt-profile fitting performed on individual systems 
%are listed in Table~1 and fits
%the one toward Q~0347$-$383 
Fits are shown in Figs.~8, 9 and 10. The fits to the absorption lines in the $z_{\rm abs}$~=~3.025 
system towards Q~0347$-$383 can be seen in Ledoux et al. (2003). 
Errors for individual components are the 1~$\sigma$ errors from Voigt 
profile fitting as given by FitLyman.
\par\noindent
Following the usual procedure, [N/H] and [O/H] abundances
are obtained by adding the column densities in all
detected individual components.
%
%For the [N/H] and [O/H] abundances, column densities of N~{\sc i} and
%O~{\sc i} are added, as usual, over all detected components.
%When the system has multiple components, we add the column densities of all O~{\sc i} components
%to derive the Oxygen abundance. 
For O~{\sc i}, all the systems are dominated by one strong 
component 
%It is possible that in some of the weak components the gas 
%is partially ionized and that some ionization correction should be applied
%to these components. However, this ionization correction is probably smaller 
%than 0.1 dex in the weakest components. Since these components 
%do not account, in total, for more than $\sim$20\% of the total column density, 
%the corresponding error 
%
%we believe we make less error by taking the weak components into account rather than 
%using the main component only. 
in which we do not expect much ionization correction 
(see next Section for details on each of the systems).
It is possible that in some of the weak components the gas 
is partially ionized and that some ionization correction should be applied
to these components. However, this ionization correction is probably smaller 
than 0.2 dex in the weakest components and these components 
do not account, in total, for more than $\sim$20\% of the total column density.
We therefore do believe that the error on [O/H] due to the procedure is less than
0.2~dex for the whole system (see also the discussion in Prochaska et al. 2002).
For N~{\sc i}, the above error is less as the transitions are weaker and the
species is always detected in less components.
\par\noindent
For the [N/O] ratio, we use the column densities in the components where
{\sl both} species are detected.
%when N~{\sc i} is detected in only one component we take 
%the ratio in this component only. 
Note that when N~{\sc i} is detected in only one component, this is always the 
strongest O~{\sc i} component. It is interesting that, in case 
N~{\sc i} and O~{\sc i} are detected in several components, as toward Q~2332$-$094, 
the [N/O] ratios are similar in all components.
\par\noindent
%
% and metallicities are derived for the systems.
We adopt the Oxygen and Nitrogen solar abundances from Lodders (2003):
12~+~(O/H)$_{\odot}$~=~8.69 and 12~+~(N/H)$_{\odot}$~=~7.83.
Note that the main discussion on the production of these elements is
little dependent on the exact adopted values. 
Depletion onto dust-grains is known to be much smaller in DLA systems
compared to the ISM of our Galaxy mostly because of smaller 
(by a factor of at least ten) metallicities. Depletions should
be therefore much smaller than in the diffuse galactic ISM clouds
where it has been shown that, apart from local effects, the [N/O] ratio
is similar to the solar ratio (Knauth et al. 2006). 
\par\noindent
  \begin{table*}
      \caption[]{Metal abundances in Damped Lyman-$\alpha$ systems}
         \label{table2}
          \centering
          \begin{tabular}{l c c c c c c c c}
          \hline\hline
Name &  $z_{\rm abs}$ & [Fe/H] & [O/H] & [N/H] & (O/H)+12$^c$ & (N/O)$^d$ & [S/H] & [Si/H] \\
\hline  
Q 0000$-$263$^a$  & 3.390  & $-$2.00(0.09) & $-$1.67(0.13) &   $-$2.50(0.08) & 7.02 &  $-$1.69(0.10)     & $-$1.89(0.09) & $-$1.88(0.08) \\
Q 0102$-$190      & 2.926  & $-$1.68(0.10) & $-$1.56(0.13) &   $-$3.07(0.15) & 7.13 &  $-$2.30(0.16) & $-$1.37(0.10) & $-$1.50(0.10) \\
Q 0112$-$306      & 2.418  & $-$2.64(0.09) & $-$2.24(0.12) &   $-$3.17(0.09) & 6.45 &  $-$1.79(0.09)     & $-$1.25(0.08) & $-$2.42(0.08) \\
Q 0347$-$383      & 3.025  & $-$1.85(0.05) & $-$0.97(0.12) &   $-$1.92(0.05) & 7.72 &  $-$1.81(0.11)     & $-$1.18(0.06) & $-$1.49(0.06) \\
Q 0841$+$129      & 2.476  & $-$1.79(0.10) & $-$1.40(0.10) &   $-$2.69(0.10) & 7.29 &  $-$2.15(0.04)     & $-$1.55(0.10) & $-$1.58(0.10) \\
Q 0913$+$072      & 2.618  & $-$2.72(0.10) & $-$2.45(0.10) &  $<-$3.73       & 6.24 &  $<-$2.14          & $-$1.66(0.10) & $-$2.57(0.10) \\
Q 1108$-$077      & 3.608  & $-$1.96(0.07) & $-$1.69(0.08) &  $<-$3.36       & 7.00 &  $<-$2.53          &               & $-$1.57(0.07) \\
Q 1337$+$113      & 2.508  & $-$2.23(0.05) & $-$1.91(0.11) &   $-$3.11(0.11) & 6.78 &  $-$2.06(0.14)     &               & $-$1.84(0.06) \\
Q 1337$+$113      & 2.796  & $-$2.14(0.08) & $-$1.95(0.10) &   $-$2.84(0.09) & 6.74 &  $-$1.61(0.09) &               & $-$1.84(0.09) \\
Q 1340$-$136      & 3.118  & $-$1.59(0.08) & $-$1.22(0.08) &   $-$2.60(0.08) & 7.47 &  $-$2.24(0.03)     & $-$1.41(0.08) & $-$1.18(0.08) \\
Q 1409$+$095      & 2.456  & $-$2.43(0.08) & $-$1.89(0.08) &  $<-$2.85       & 6.80 &  $<-$1.82          & $-$1.74(0.08) & $-$2.17(0.08) \\
Q 1409$+$095      & 2.668  & $-$1.51(0.08) & $-$1.31(0.08) &   $-$2.13(0.08) & 7.38 &  $-$1.68(0.03)     & $-$1.40(0.09) & $-$1.28(0.08) \\
Q 1946$+$769$^b$  & 2.844  & $-$2.50(0.06) & $-$2.14(0.06) &   $-$3.51(0.07) & 6.55 &  $-$2.23(0.04)     &               & $-$2.21(0.06) \\
Q 2059$-$360      & 3.083  & $-$1.97(0.08) & $-$1.58(0.09) &   $-$2.86(0.08) & 7.11 &  $-$2.14(0.04)     & $-$1.76(0.09) & $-$1.66(0.09) \\
Q 2332$-$094      & 3.057  & $-$1.60(0.07) & $-$1.24(0.07) &   $-$2.60(0.08) & 7.45 &  $-$1.94(0.05) & $-$1.32(0.08) &               \\
QXO 0001$^b$      & 3      &$<-$1.08       & $-$1.62(0.05) &   $-$3.22(0.06) & 7.07 &  $-$2.46(0.05)     &               & $-$1.79(0.05)  \\
       \hline
\multicolumn{9}{l}{$^a$Molaro et al. (2001); $^b$Prochaska et al. (2002); $^c$$\sigma$~=~$\sigma$([O/H])}\\
\multicolumn{9}{l}{$^d$Errors are taken as square root of the geometrical mean of the variances of $N$(N~{\sc i}) and $N$(O~{\sc i}) as given in}\\
\multicolumn{9}{l}{Table~1 except for Q 0102$-$190, Q 1337$+$113 at $z_{\rm abs}$~=~2.796 and  Q 2332$-$094 for which we consider only}\\
\multicolumn{9}{l}{the main components of the systems to calculate the [N/O] ratio (see Text in Section 3).}
         \end{tabular}
%      \begin{list}{}{}
%     \item[$^{\mathrm{a}}$] Molaro et al. (2000)
%%     \item[$^{\mathrm{b}}$] also Ledoux et al. (2003)
%%\item[$^{\rm c}$] also Erni et al. (2006)
%%\item[$^{\rm d}$] $3\sigma$ detection limit.
%%\item[$^{\rm e}$] also Pettini et al. (2002)
%\item[$^{\rm b}$] Prochaska \& Wolfe (2002)
%     \end{list}
  \end{table*}

\section{Comments on individual systems}
Q 0000$-$263: We adopt the column densities from Molaro et al. (2001). The Oxygen abundance 
is derived from unsaturated O~{\sc i}$\lambda$$\lambda$925,950 absorption lines.
The Nitrogen abundance is derived from the unsaturated $\lambda$953 line and the
$\lambda$1134 triplet. The system is modeled with only one component.
\par\noindent 
Q 0102$-$190 (Fig.~8, top panel): The structure of the system is derived from the fit of Fe~{\sc ii} 
and Si~{\sc ii} lines. $N$(N~{\sc i}) is derived from a $\lambda$1199 optically thin feature 
corresponding in redshift with the strongest O~{\sc i} component. 
We add all O~{\sc i} column densities (obtained from optically thin
$\lambda\lambda$976,1039 features) to calculate the O abundance. Note however
that the red satellites do not contribute much. The [N/O] ratio is taken as the
column density ratio in the main central component.
\par\noindent
Q 0112$-$306 (Fig.~8): The two-component structure of the system is very well defined
from Si~{\sc ii} and Fe~{\sc ii} absorption lines. This is why we are confident
that the log~$N$(O~{\sc i}) column density derived from the moderately saturated
$\lambda$1302 absorption is
robust. N~{\sc i} is detected through an optically thin $\lambda$1199 feature.
\par\noindent
Q 0347$-$383: This system has been studied by Levshakov et al. (2002) and
Prochaska et al. (2002). The UVES data have also been reanalysed by Ledoux 
et al. (2003). $N$(N~{\sc i}) is constrained by optically thin $\lambda$1134 
features and $N$(O~{\sc i}) by $\lambda$950 and $\lambda$974. Note that
Levshakov et al. (2002) derived log~$N$(N~{\sc i})~=~14.89, 0.25~dex larger than
our result. We believe this is because the latter authors do not restrict their
fit to optically thin lines.
\par\noindent
Q 0841+129 (Fig.~8): This is a beautiful one-component system with several
optically thin N~{\sc i} components (see also Centuri\'on et al. 2002). 
$N$(O~{\sc i}) is derived from 
the consistent fit of optically thick $\lambda$1302 and 1039 features
and the moderately saturated $\lambda$950 absorption.
Centuri\'on et al. (2003) found a $N$~{\sc i} column density 0.16~dex
larger. This is likely due to the difference in data quality.
\par\noindent
Q 0913+072 (Fig.~8, bottom panel): The absorption profile is made of two
closely blended components of similar strength.
%This system has two blended similarly strong components.
An upper limit on log~$N$(N~{\sc i}) is obtained from $\lambda$1199
(a feature is present but below the 3$\sigma$ detection limit, see also
Erni et al. 2006). Upper limits are derived from the noise
in the adjacent continuum; when two components are present we conservatively
consider each component independently.
$N$(O~{\sc i}) is derived from optically thin $\lambda$1039.
\par\noindent
Q 1108$-$077 (Fig.~9, top panel):  This is a one-component system.
An upper limit on $N$(N~{\sc i}) is obtained from $\lambda$1134. 
$N$(O~{\sc i}) is derived from an optically thin $\lambda$1039 feature.
\par\noindent
Q 1337+113, $z_{\rm abs}$~=~2.508 (Fig. 9): This is again a single component
system with well constrained N~{\sc i} and O~{\sc i}
column densities from optically thin, respectively, $\lambda$1200 and $\lambda$1039
absorption lines.
\par\noindent
Q 1337+113, $z_{\rm abs}$~=~2.796 (Fig.~9): There are two components 
in this system. $N$(N~{\sc i}) and $N$(O~{\sc i}) are both well defined
from optically thin, respectively, $\lambda$1134,1200 and $\lambda$976 features.
The component at $z_{\rm abs}$~=~2.79581 is the strongest and is the
only component detected in N~{\sc i}. It is possible that N~{\sc i} is
somewhat ionized in the component at $z_{\rm abs}$~=~2.79557. 
As done for all systems, we therefore add the O~{\sc i} 
column densities to calculate the Oxygen abundance but take the N/O ratio from
the component at $z_{\rm abs}$~=~2.79581.
\par\noindent
Q 1340$-$136 (Fig.~9, bottom panel): This is a beautiful one-component system with optically
thin transitions.
\par\noindent
Q 1409+095, $z_{\rm abs}$~=~2.456 (Fig.~10, top panel): Although there may be
some absorption at the wavelengths corresponding to the
N~{\sc i} transitions, we derive only an upper limit on $N$(N~{\sc i})
in this two-component system (see also Pettini et al. 2002).
The O~{\sc i} column density is well defined by optically thin
$\lambda$1039 absorption.
\par\noindent
Q 1409+095, $z_{\rm abs}$~=~2.668 (Fig.~10): The consistency of
the $\lambda\lambda$1199,1200 and $\lambda$1134 features makes us
believe that N~{\sc i} is present in this one-component system.
The O~{\sc i} column density is well defined by optically thin
$\lambda$1039 and $\lambda$950 absorptions.
\par\noindent
Q 1946+769: We use the measurements by Prochaska et al. (2002).
\par\noindent
Q 2059$-$360 (Fig.~10): N~{\sc i} and O~{\sc i} are detected in
the three components of this system in, respectively, optically thin
$\lambda$1134,1200 and $\lambda$950,976 transitions. It is important 
to note that the [N/O] ratio is similar for the three components.
This supports the assumption made in this paper that ionization 
corrections are negligible.
\par\noindent
Q 2332$-$094 (Fig.~10, bottom panel): The system is complex but dominated by two main
components. We add all O~{\sc i} column densities but the contribution of the 
three satellite components is negligible. N~{\sc i} is detected in a single component 
corresponding to the strongest O~{\sc i} component at $z_{\rm abs}$~=~3.05723
and we use the [N/O] ratio in this component. 
\par\noindent
QXO 0001: We use the measurements by Prochaska et al. (2002). Note that
only one digit is given by these authors for the absorption redshift. 
\section{Results}
%
%Plot the histogramme of [N/O] => dichotomie.
%
A summary of the abundances of interest here, measured in
the sixteen systems,
 is given in Table~2.
\subsection{The Oxygen abundance in DLAs}
\begin{figure}
   \centering
   \includegraphics[width=7.5cm,angle=-90,clip]{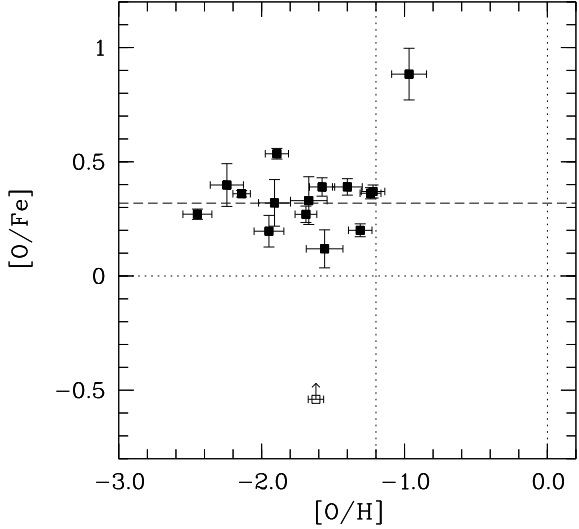}
   \caption{[O/Fe] abundance ratio versus Oxygen abundance [O/H].
The dashed and dotted lines are drawn for convenience.
%indicates the standard definition of DLA systems (log~$N$(H~{\sc i})~$>$~20.3).
\label{OFe_OH}} 
    \end{figure}
Few Oxygen abundance measurements have been reported up to now in DLAs
(see Molaro et al. 2000; Pettini et al. 2002).
The main reason is that most of the time O~{\sc i}$\lambda$1302 absorption
lines are badly blended or saturated. High resolution and high SNR spectra covering the bluer
part of the optimal wavelength range are needed to detect weaker O~{\sc i} transitions. 
%feature corresponds to the strong $\lambda$1302 transition
%that is often badly saturated. 
Thanks to the large number of DLAs observed in the course of
the survey for molecular hydrogen and to the UVES sensitivity into the blue
we could built a sample of robust measurements.

\begin{figure}
   \centering
   \includegraphics[width=7.5cm,angle=-90,clip]{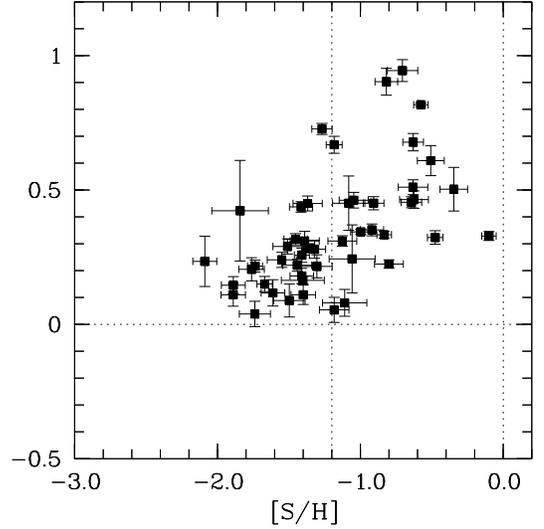}
   \caption{Sulfur to Iron abundance ratio, [S/Fe], versus Sulfur 
abundance relative to solar, [S/H]. The lines are drawn for 
convenience.
\label{depletion}} 
    \end{figure}

Fig.~\ref{OFe_OH} gives the [O/Fe] abundance ratio versus the Oxygen abundance.
In principle, this ratio should be corrected for depletion of metals onto
dust-grains. However this correction is known to be negligible for [X/H]~$<$~$-$1.2
(Prochaska \& Wolfe 2003; Ledoux et al. 2003; Wolfe et al. 2005). 
To support this idea, we plot in Fig.~\ref{depletion} the abundance 
ratio [S/Fe] versus the Sulfur abundance [S/H] for all systems in our DLA sample
(Ledoux et al. 2006a). 
Sulfur and Iron are known to be, respectively, little and strongly depleted onto 
dust-grains. It is apparent that the scatter of the [S/Fe] values is much less 
below [S/H]~=~$-$1.2 than above. This behavior is a consequence of the 
largest depletion of Iron onto dust-grains for [S/H]~$>$~$-$1.2.

It can be seen that because of our choice to reject the systems where all 
O~{\sc i} transitions are badly saturated,
our sample is biased against metallicities [O/H]~$>$~$-$1. All
our systems except one have metallicities in the range $-$2.5~$<$~[O/H]~$<$~$-$1.2.
The stricking result here is that the [O/Fe] ratio is pretty constant over
the range studied. The mean value of the 14 measurements
at [O/H]~$<$~$-$1 (therefore excluding Q~0347$-$383)
is [O/Fe]~=~0.32$\pm$0.10 with a slight tendency
of this ratio to increase at lower metallicity. The only departure
from this picture is the point at [O/H]~=~$-$0.97, [O/Fe]~=~0.88, 
from the system at $z_{\rm abs}$~=~3.025 towards Q~0347$-$383. This is
probably due to depletion onto dust-grains ([Zn/Fe]~=~0.72).
Actually, molecular hydrogen, although with very low molecular fraction, 
is detected in this system (Levshakov et al. 2002, Ledoux et al. 2003),
supporting the presence of dust. 

\begin{figure}
   \centering
   \includegraphics[width=7.5cm,angle=-90,clip]{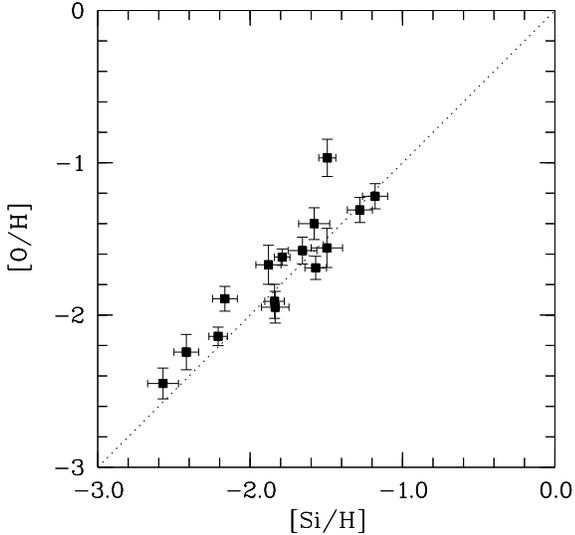}
   \caption{Oxygen abundance relative to solar, [O/H], versus 
Silicon abundance [Si/H]. Dotted line is for [O/H]~=~[Si/H].
%indicates the standard definition of DLA systems (log~$N$(H~{\sc i})~$\geq$~20.3).
\label{SiO}} 
    \end{figure}

As the Oxygen abundance is difficult to measure, the Si abundance is often used
instead. Prochaska \& Wolfe (2002) and Dessauges-Zavadsky et al.
(2006) have found that the [Si/Fe] ratio is pretty constant amongst
DLA systems around a value of +0.43 with a relatively small dispersion.
We plot in Fig.~\ref{SiO} the [O/H] abundance versus the [Si/H] abundance.
It is apparent that the two abundances correlate well. There may be
a slight tendency for Oxygen to be more abundant than Silicon for
[O/H]~$<$~$-$2 but there are only four measurements there. 
%This could however explain why the [X/Fe] ratio is slightly (0.1~dex) larger
%for Silicon than for Oxygen.

\subsection{The [N/O] abundance ratio}

Since we will use two different notations, we would like first to clarify them.
The (N/O) abundance ratio is simply the ratio of the Nitrogen to Oxygen
abundances, (O/H)~=~log$N$(O~{\sc i})$-$log$N$(H~{\sc i}) and the same for 
(N/H). This ratio is often plotted versus the Oxygen abundance written as
(O/H)~+~12. It is however very convenient, and usual, to refer to solar abundances,  
[O/H]~=~(O/H)$-$(O/H)$_{\odot}$.
In that case solar abundances should be specified. We use: 
12~+~(O/H)$_{\odot}$~=~8.69 and 12~+~(N/H)$_{\odot}$~=~7.83.
Therefore [N/O]~=~(N/O)~$+$~0.86.
What is most important in the following is the value assigned 
to the so-called "primary plateau". Nava et al. (2006) observationally
derive a primary plateau at (N/O)$_{\rm PP}$~=~$-$1.43 with objects distributed within
a range of $-$1.54 to $-$1.27. In that case, [N/O]$_{\rm PP}$~=~$-$0.57.  
From measurements in blue compact dwarf (BCD) galaxies Izotov \& Thuan (2004)
%Izotov et al. (2006) 
however find slightly smaller values with a plateau
at about (N/O)$_{\rm PP}$~=~$-$1.6 or [N/O]$_{\rm PP}$~=~$-$0.74.

\begin{figure}
   \centering
   \includegraphics[width=7.5cm,angle=-90,clip]{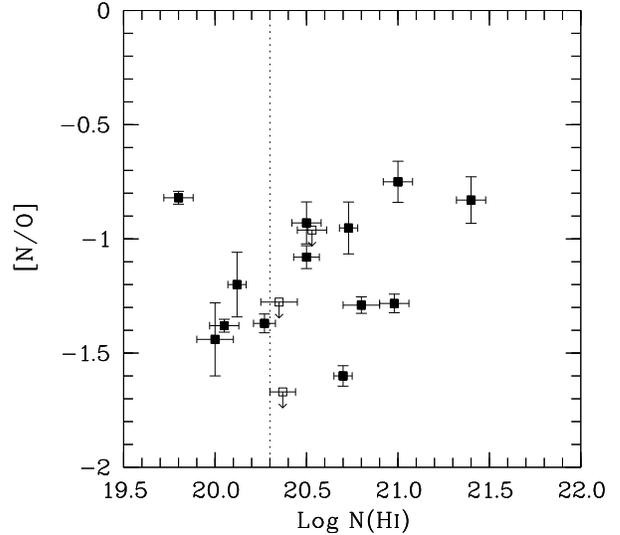}
   \caption{[N/O] abundance ratio versus neutral hydrogen column density
log~$N$(H~{\sc i}). The vertical dotted line indicates the standard definition of DLA systems
(log~$N$(H~{\sc i})~$>$~20.3).
\label{NO-HI}} 
    \end{figure}
%bb=55 80 550 755,
%
\begin{figure}
   \centering
   \includegraphics[width=6.5cm,angle=-90,clip]{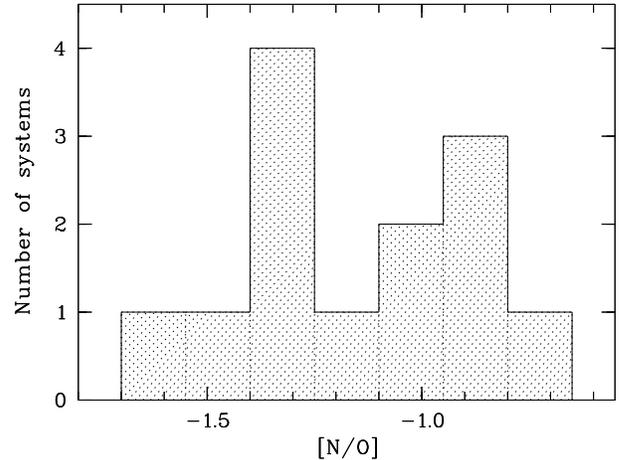}
   \caption{Histogram of [N/O] abundance ratio for the 13 systems with definite measurements
in our sample.
\label{histoNO}} 
    \end{figure}
\begin{figure}
   \centering
   \includegraphics[width=7.5cm,angle=-90,clip]{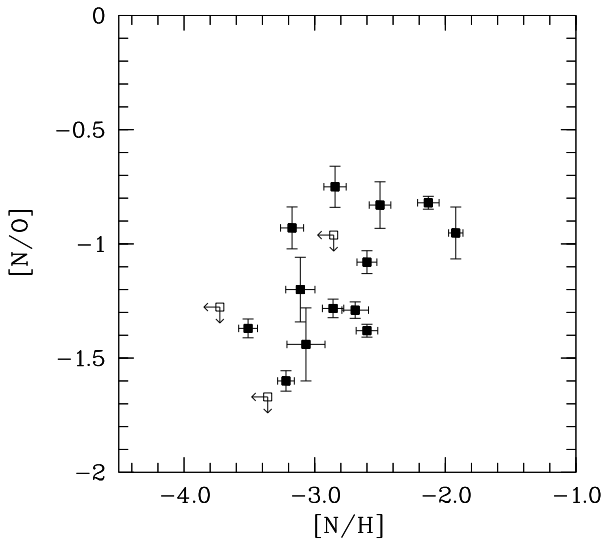}
   \caption{Nitrogen to Oxygen abundance ratio, [N/O],
versus Nitrogen abundance [N/H]. [X/H]~=~(X/H)~$-$~(X/H)$_{\odot}$.
\label{NO-NH}} 
    \end{figure}

%\begin{figure}
%   \centering
%   \includegraphics[width=6.5cm,angle=-90,clip]{graphNO_H.ps}
%   \caption{Histogramme of [N/O] abundance ratio in our sample.
%\label{NO-H}} 
%    \end{figure}

%
In Fig.~\ref{NO-HI} the [N/O] abundance ratio is plotted versus the neutral
hydrogen column density. It can be seen that there is no apparent correlation
between the two parameters. 
The smallest [N/O] value is found for a system with log~$N$(H~{\sc i})~$>$~20.3
and one of the largest values is found for the system with the lowest log~$N$(H~{\sc i}).  
This gives additional confidence in the assumption that the
ionization correction is negligible when deriving the abundances from 
$N$(O~{\sc i}), $N$(N~{\sc i}) and $N$(H~{\sc i}) (Viegas 1995).
\par
There is a possible dichotomy between systems with [N/O]~$\sim$~$-$0.9 and $-$1.4. 
%will be noticeable in all Figures. 
It can be seen on Fig.~\ref{histoNO} that the distribution of the [N/O] abundance ratio 
has two peaks, one around [N/O]~$\sim$~$-$1.35 and one around [N/O]~$\sim$~$-$0.9. 
%The effect is not strongly marked however. 
This could be however a consequence of small number statistics.
Out of 13 systems with definite measurements,
7 have [N/O]~$<$~$-$1.15 and 5 have [N/O]~$>$~$-$1.0. If we add
the two upper limits with [N/O]~$<$~$-$1.2, we find that the number of systems
with [N/O]~$<$~$-$1.15 is about twice larger than those with [N/O]~$>$~$-$1.0.
A similar dichotomy has already been noted in the [N/$\alpha$] ratio
by Centuri\'on et al. (2003). They find however that 75\% of systems
cluster around [N/$\alpha$]~$\sim$~$-$0.87 and the remaining 25\% cluster around
[N/$\alpha$]~$\sim$~$-$1.45. 
%We therefore disagree on the numbers. 
Also, note that the dichotomy present in our data is less pronounced than what is
claimed by Centuri\'on et al. (2003) because of a large dispersion of the
values below [N/O]~$\sim$~$-$1.2 (see also Henry \& Prochaska 2007). 
One could wonder if these differences could be due to our sample 
being biased against high Oxygen metallicity systems compared to the Centuri\'on 
et al. sample. We do not think this is the case because there is no obvious
correlation between [O/H] and [N/O].
\par
In addition, Centuri\'on et al. (2003) claimed that there is a transition between the two 
regimes at [N/H]~$\sim$~$-$2.8 with most, if not all, of the [N/$\alpha$]~$<$~$-$1.5 
systems having [N/H]~$<$~$-$2.8. Molaro et al. (2003) place this transition
at [N/H]~$\sim$~$-$3.0 and argue that this could be understood as the point at
which Nitrogen production from AGB stars begins to dominate that of massive stars.
%It can be seen on Fig~\ref{NO-NH} that this is not the case. 
Although all the points with [N/O]~$\leq$~$-$1.2 have 
[N/H]~$<$~$-$2.4, we find two systems with [N/O]~$>$~$-$1.2 and
[N/H]~$\sim$~$-$3.0. Again, below [N/O]~$<$~$-$1, the scatter in the measurements
is larger than above this limit. 
%The [N/O] vs [N/H] plot resembles
%the usual [N/O] vs [O/H] plot we will discuss in the following.
Note however that the four points with [N/H]~$<$~$-$3.2 have [N/O]~$<$~$-$1.4 
so that the figure is not inconsistent with some kind of break below [N/H]~=~$-$3.2.
More data are needed below this limit before any firm conclusion can be drawn.
%Note however that the point at [N/H]=-3.15 has [N/O]=-0.9.

%The dichotomy is better seen on Fig.~\ref{NO-H} where the [N/O] abundance ratio is plotted versus
%the total neutral hydrogen column density, log~$N$(H~{\sc i}). The figure
%clearly shows as well that there is no correlation between the two
%parameters.

\begin{figure}
   \centering
\vbox{ 
   \includegraphics[width=7.5cm,angle=-90,clip]{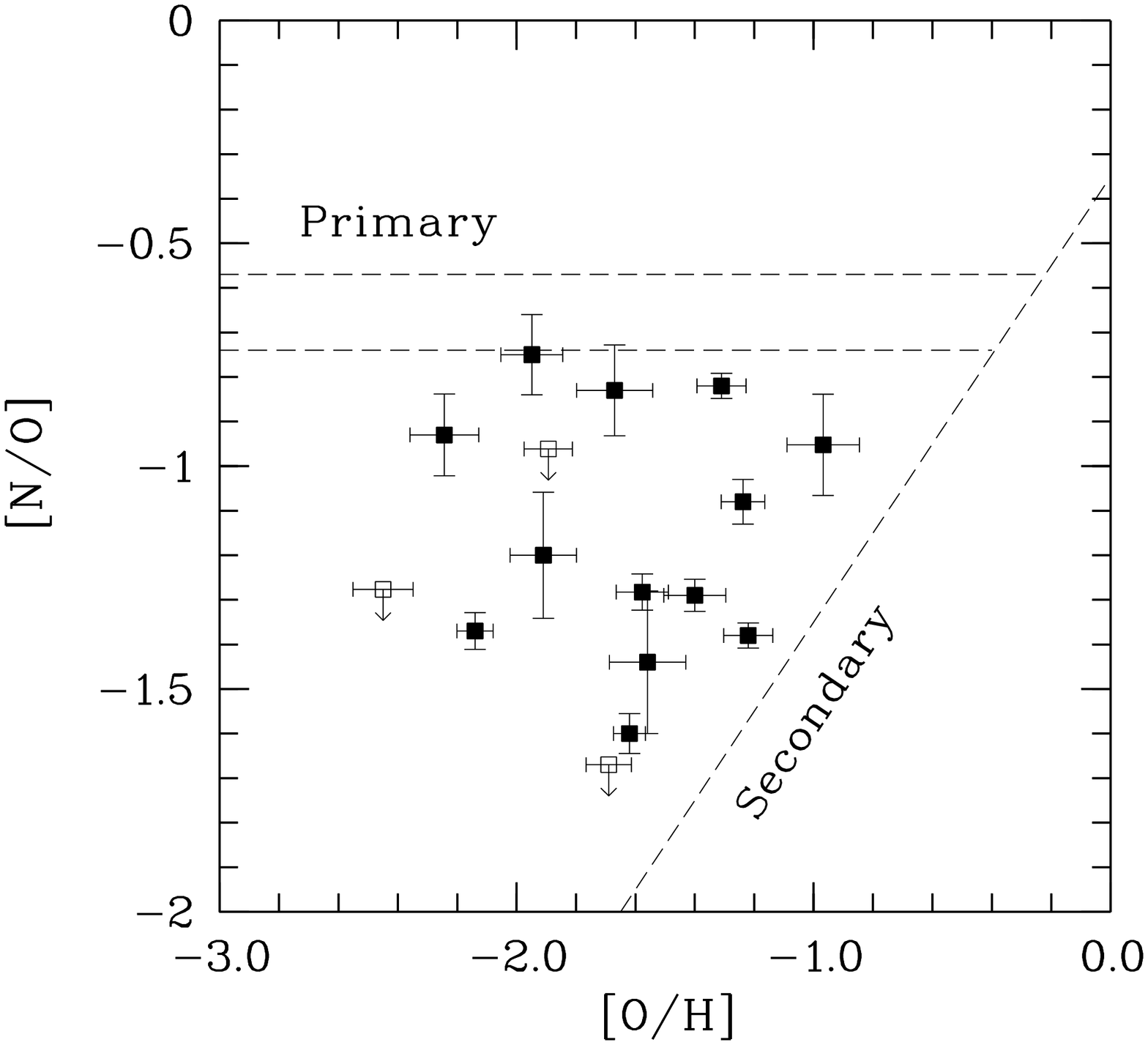}
   \includegraphics[width=7.5cm,angle=-90,clip]{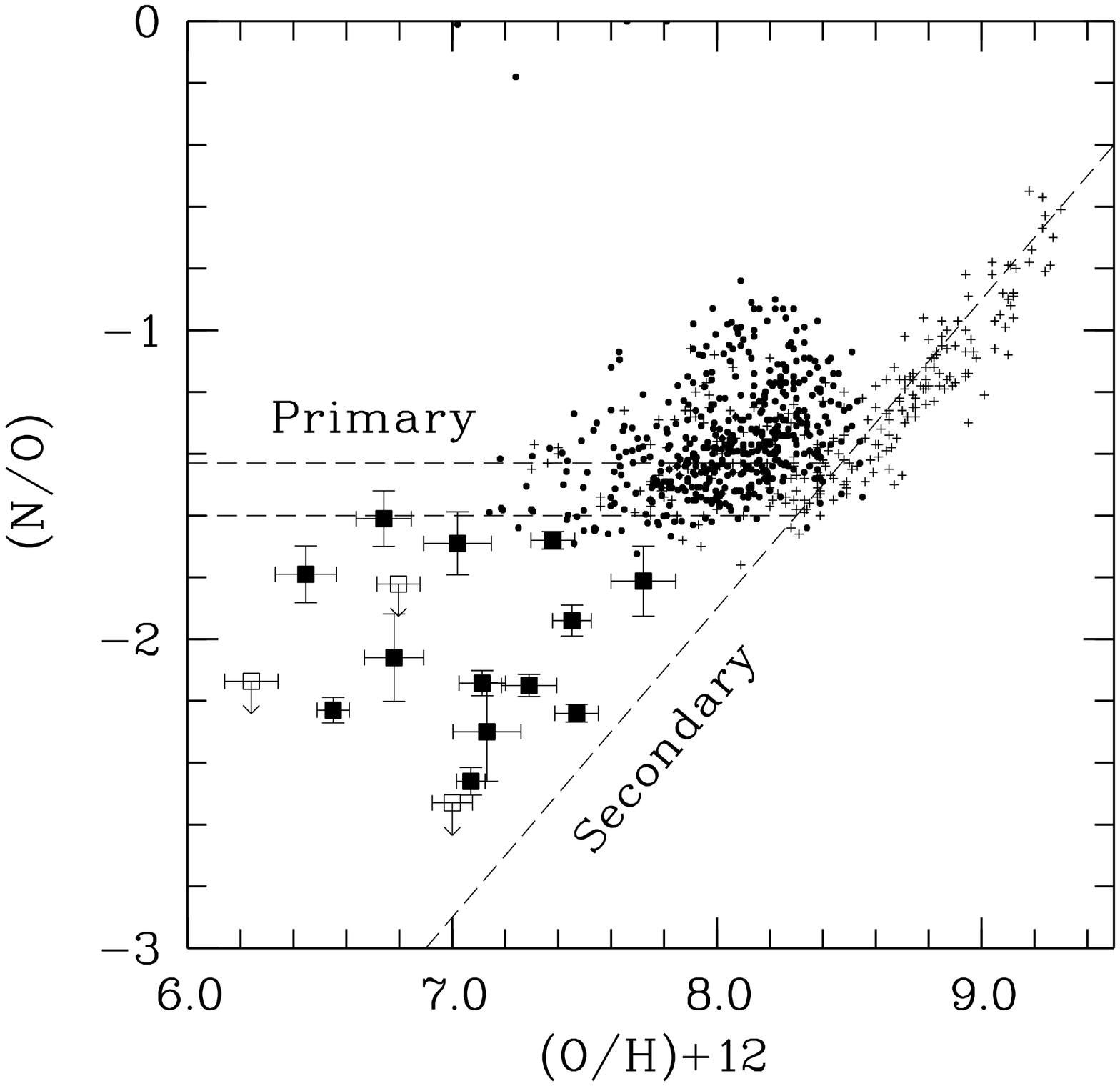}
}
   \caption{{\sl Upper panel}: Nitrogen to Oxygen abundance ratio, [N/O],
versus Oxygen abundance [O/H]. [X/H]~=~(X/H)$-$(X/H)$_{\odot}$. 
The range for the local primary plateau 
is indicated as two horizontal dashed lines at [N/O]~=~$-$0.57 and $-$0.74. 
The other dashed line indicates the expected position in case of secondary production:
the line is the extrapolation at low [N/O] values of local
measurements. {\sl Lower panel}: Same as above using the absolute 
abundances (X/H)~=~log~$N$(X)~$-$~log~$N$(H) with X~=~O or N.
The dots and crosses indicate local measurements from dwarf-irregular galaxy HII regions 
(crosses: van Zee \& Haynes 2006), HII regions in spiral galaxies (crosses: van Zee et al. 1998),
metal-poor emission line galaxies (dots, Nava et al. 2006), HII regions in
Blue Compact Dwarf galaxies (Izotov \& Thuan 2004) and emission-line galaxies
from SDSS (dots, Izotov et al. 2006).
%from SDSS (crosses, Izotov et al. 2006).
%(crosses, Nava et al. 2006, Izotov \& Thuan 2004). 
\label{NO-OH}} 
    \end{figure}

The classical plot giving [N/O] versus [O/H] is shown in Fig.~\ref{NO-OH}.
Several features have to be noted here. First, all the DLA measurements
are located in the region delineated by the usual primary and secondary 
lines. Recall that, by definition, if Nitrogen is a primary element, it should be
produced at the same time as Oxygen and the [N/O] ratio should be a constant
whatever the Oxygen metallicity is. The primary plateau observed locally
is derived from measurements of abundances in nearby low-metallicity 
galaxies: [N/O]$_{\rm PP}$ in the range $-$0.57 to $-$0.74.
If Nitrogen is a secondary element, its
production is favored at large Oxygen abundance and the [N/O] ratio
should increase with increasing Oxygen abundance.
The fact that all the DLA systems are found in this region may indicate
that at least part of the systems are in the transition zone as a consequence
of delays in the release of heavy elements from intermediate mass stars. 
This is what would be expected if during a starburst, 
at the beginning, high mass stars eject material with low [N/O] ratio 
and, after some time, intermediate mass stars eject material with higher
ratio. The pathway of one particular system in the diagram during his lifetime 
could be a line starting from the left-bottom corner of the Figure towards the
up-right direction.

Secondly, DLA measurements
are all below the local primary plateau ([N/O]$_{\rm PP}$~$\sim$~$-$0.57 to
$-$0.74). If we believe that the five top-most [N/O] DLA measurements define 
what can be considered as a plateau, this plateau is at [N/O]~$\sim$~$-$0.9, 
therefore at least $-$0.15~dex below the local primary plateau.
Note that this has not been recognized by studies based on [N/Si] or
[N/S] ratios (Centuri\'on et al. 2002, Pettini et al. 2007). 
We understand this conclusion is based on small number statistics and should
be confirmed with more data. One possibility to explain this would be
that DLAs reach the primary plateau only for [O/H]~$>$~$-$1.

%The two plateaux could be reconciled however if we correct for an Oxygen 
%overabundance as indicated by the [O/Fe] ratio of a factor of $\sim$2 (0.3 dex). 
%In that case the DLA points should be shifted
%by 0.3~dex upwards (and then fall on top of the local primary plateau) and 
%to the left. 

In the lower panel of Fig.~\ref{NO-OH}, we plot 
(N/O) vs (O/H)~+~12 for DLAs (squares) and local measurements
(dots and crosses). The latters are from dwarf-irregular galaxy HII regions (van Zee 
\& Haynes 2006), HII regions in spiral galaxies (van Zee et al. 1998),
metal-poor emission line galaxies (Nava et al. 2006; Izotov et al. 2006). 
The same remark as above is of course to be made here.
It is apparent that DLA measurements are all below the local
measurements. If a plateau is to be seen in DLA measurements, this is 
at (N/O)~$\sim$~$-$1.75. 
%This could probably be interpreted as the plateau
%of primary production shifted below the local plateau at $-$1.43. 
%because of the over-abundance of Oxygen.
%Nava et al. (2006) find that the range in log~(O/N) is of $-$1.54 to $-$1.27 in low-metallicity
%emission line galaxies.  
Other points are pretty well scattered in the diagram and we find difficult to confirm the
presence of a second plateau in Fig.~\ref{NO-OH} as claimed by 
Centuri\'on et al. (2003). 
%have claimed the presence of second plateau at [N/O]~$\sim$~$-$1.4 or (O/N)~$\sim$~$-$2.2. 
%Although the [N/O] distribution shows a two-peak feature, we find difficult to confirm the
%presence of a second plateau in Fig.~\ref{NO-OH}.
%On the contrary, one could say that the points seem to follow the so-called secondary frontier.

%La premiere version de la figure (hii_1.ps) contient les meilleures
%references, recentes, que j'ai trouvees:
%van Zee \& Haynes 2006ApJ...636..214V (dwarf irregular galaxy HII regions)
%van Zee et al. 1998AJ....116.2805V (HII regions in spiral galaxies)
%Nava et al. 2006ApJ...645.1076N
%Izotov et al. 2006A&A...448..955I
%Kobulnicky \& Skillman (1996), cf. Pettini et al. (2002) (cet
%echantillon contient des regions HII de faibles metallicites). 

% Spectres

\begin{figure}
   \centering
\vbox{ 
   \includegraphics[width=7.5cm,angle=0,clip]{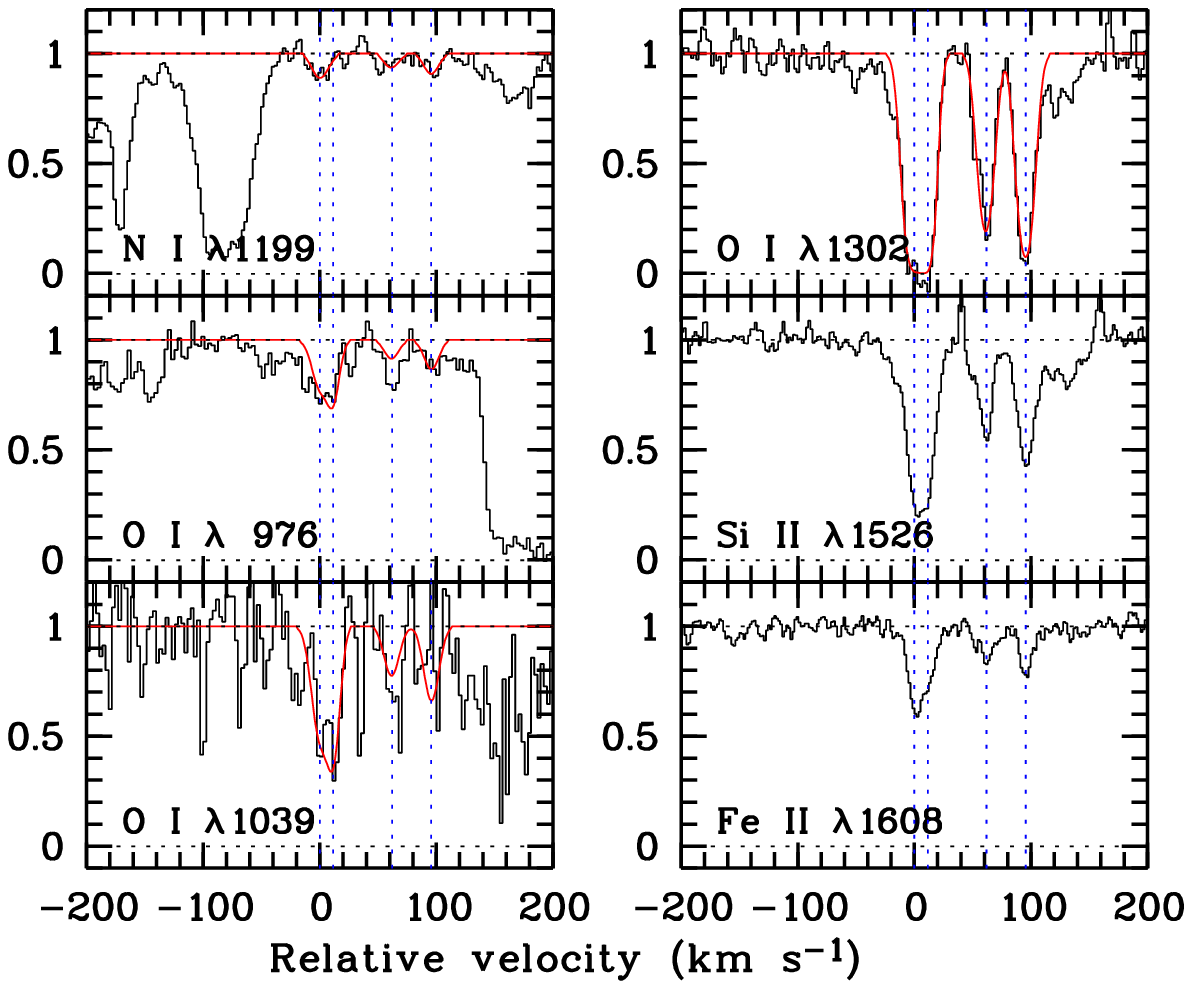}
   \includegraphics[width=7.5cm,angle=0,clip]{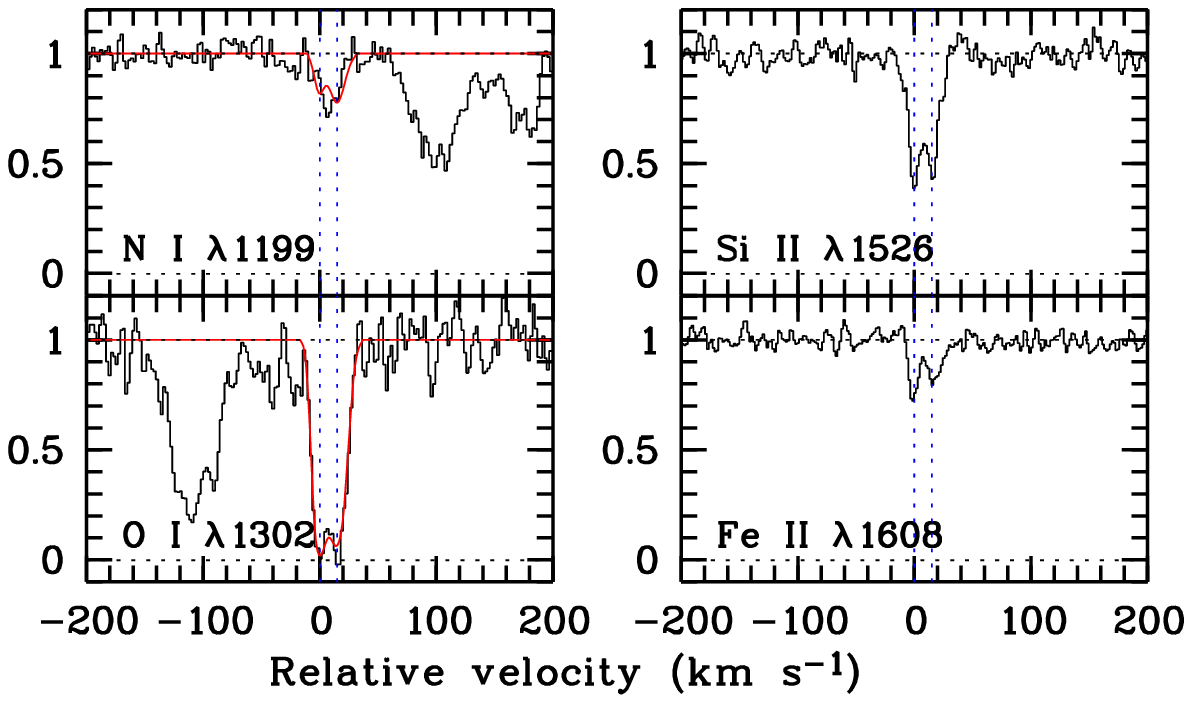}
   \includegraphics[width=7.5cm,angle=0,clip]{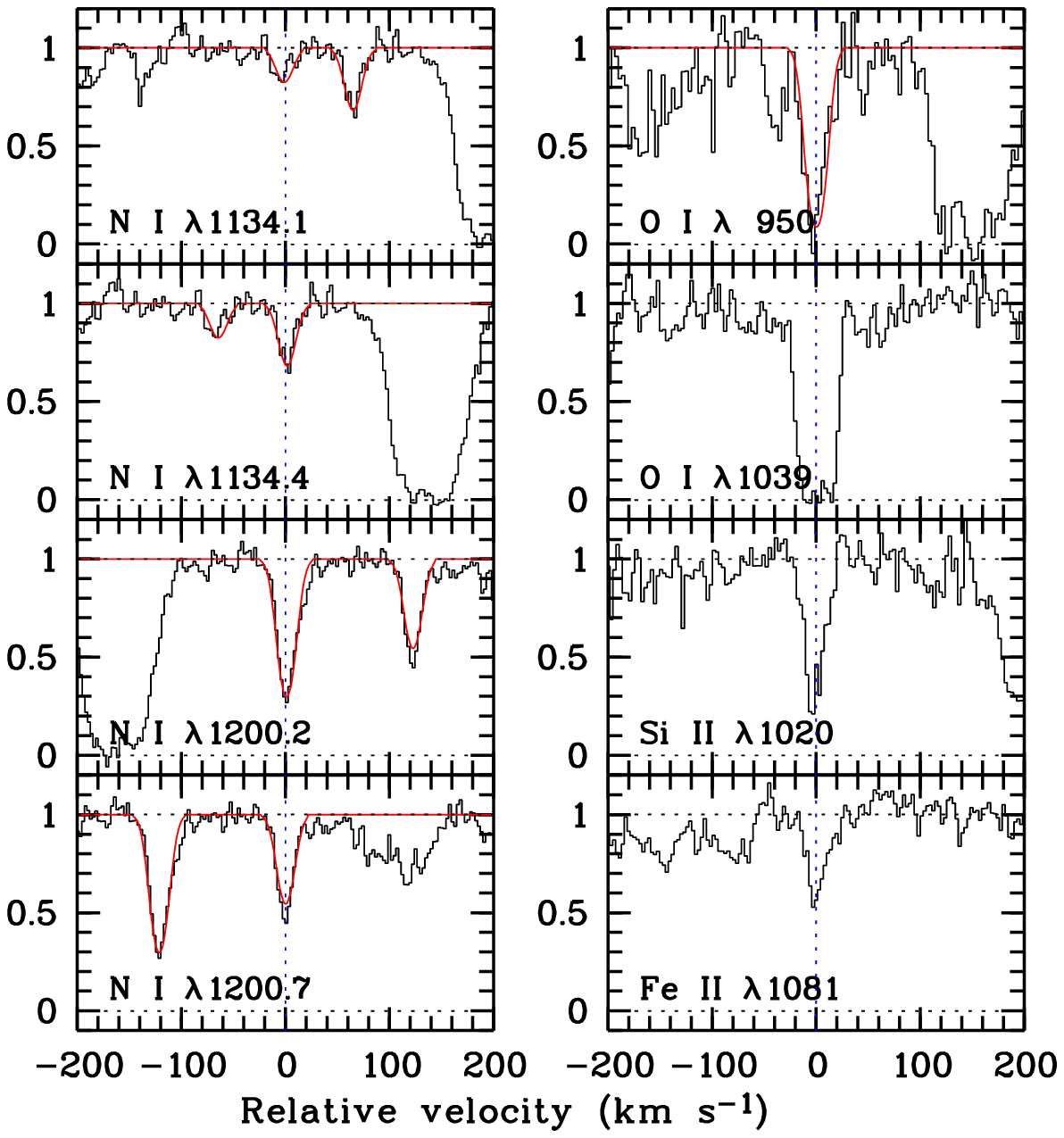}
   \includegraphics[width=7.5cm,angle=0,clip]{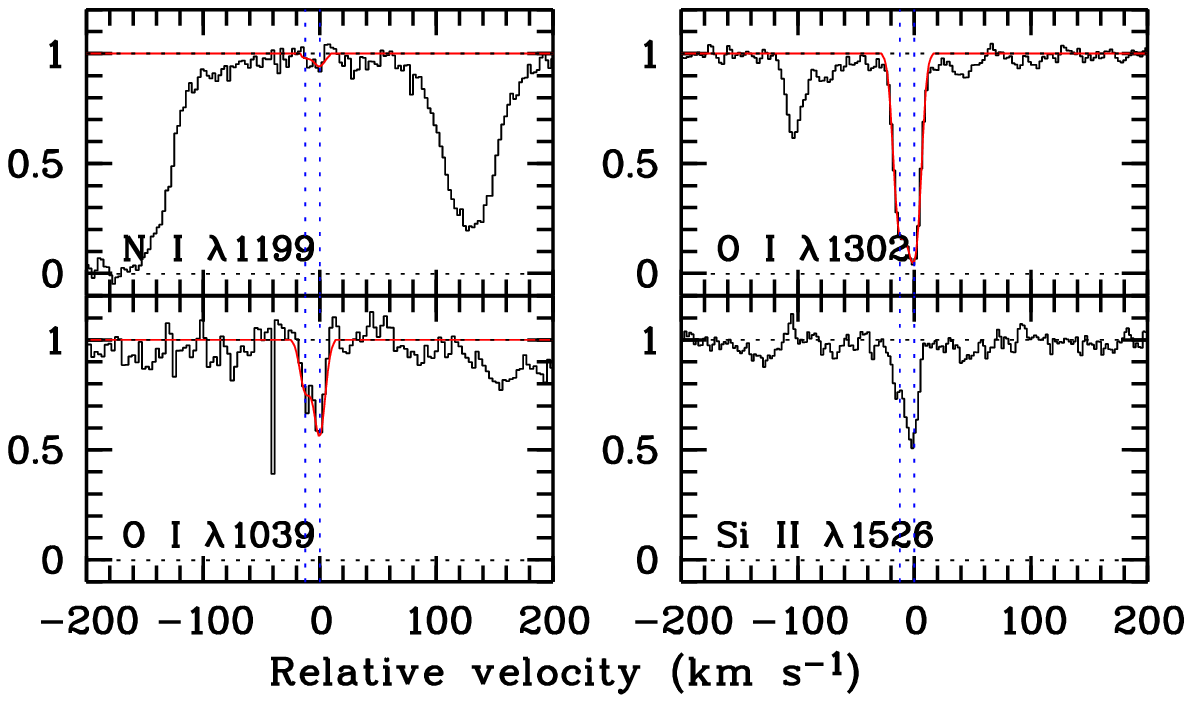}
}
   \caption{N~{\sc i} and O~{\sc i} absorption lines in the DLA systems:
Q~0102$-$190, $z$~=~2.926; Q~0112$-$306, $z$~=~2.418;
Q~0841$+$129, $z$~=~2.476 and Q~0913$+$072, $z$~=~2.618 
from top to bottom respectively. 
Model fits are overplotted.
\label{Spectres1}} 
    \end{figure}
\begin{figure}
   \centering
\vbox{ 
 \includegraphics[width=7.5cm,angle=0,clip]{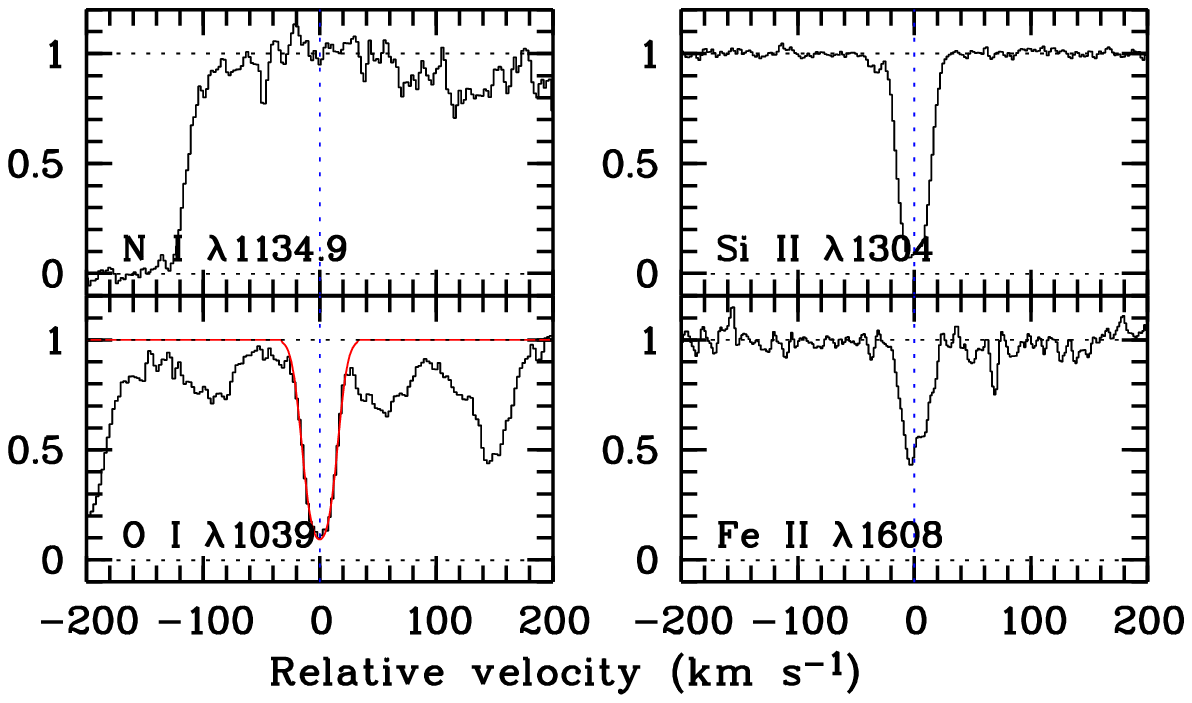}
   \includegraphics[width=7.5cm,angle=0,clip]{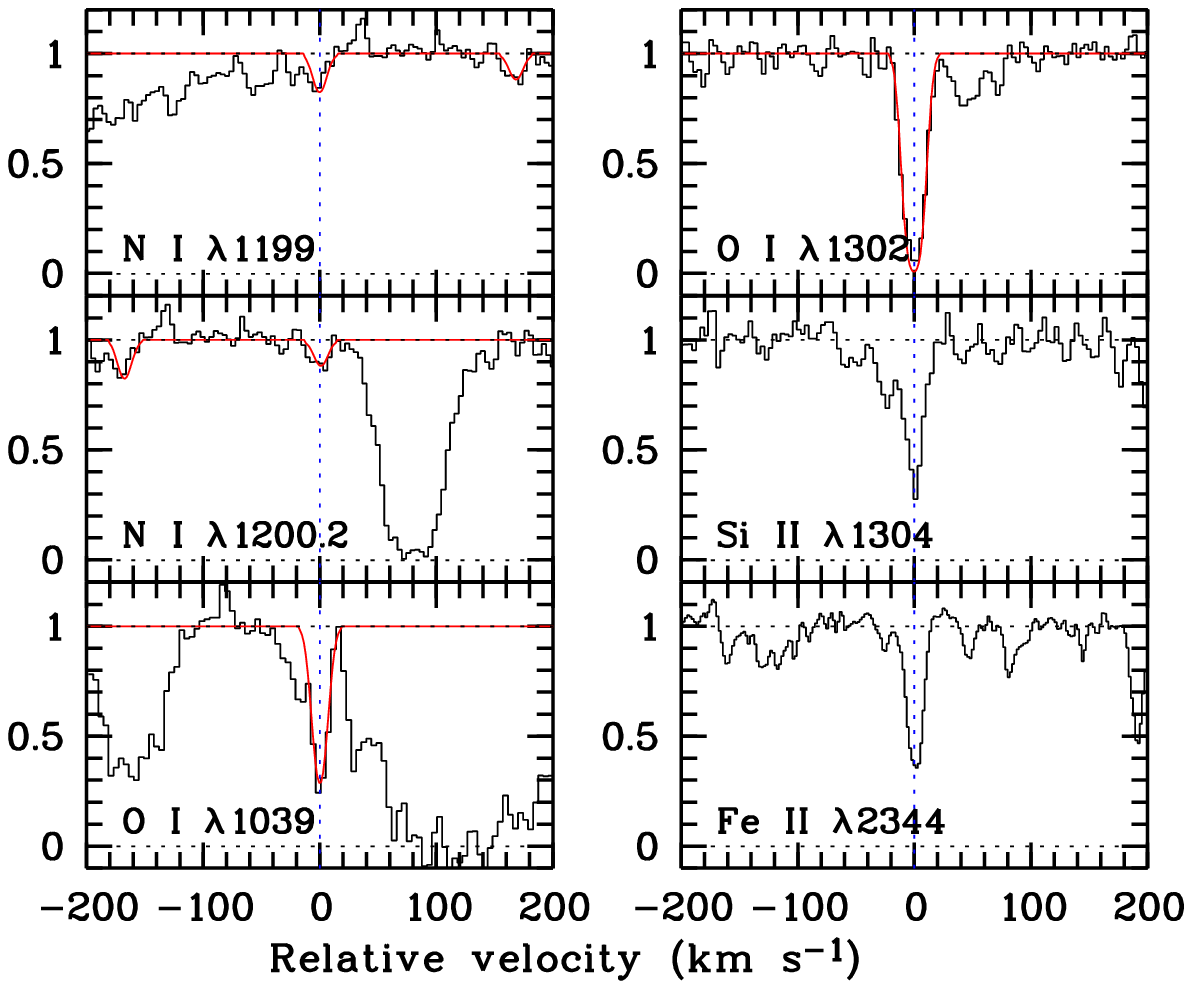}
   \includegraphics[width=7.5cm,angle=0,clip]{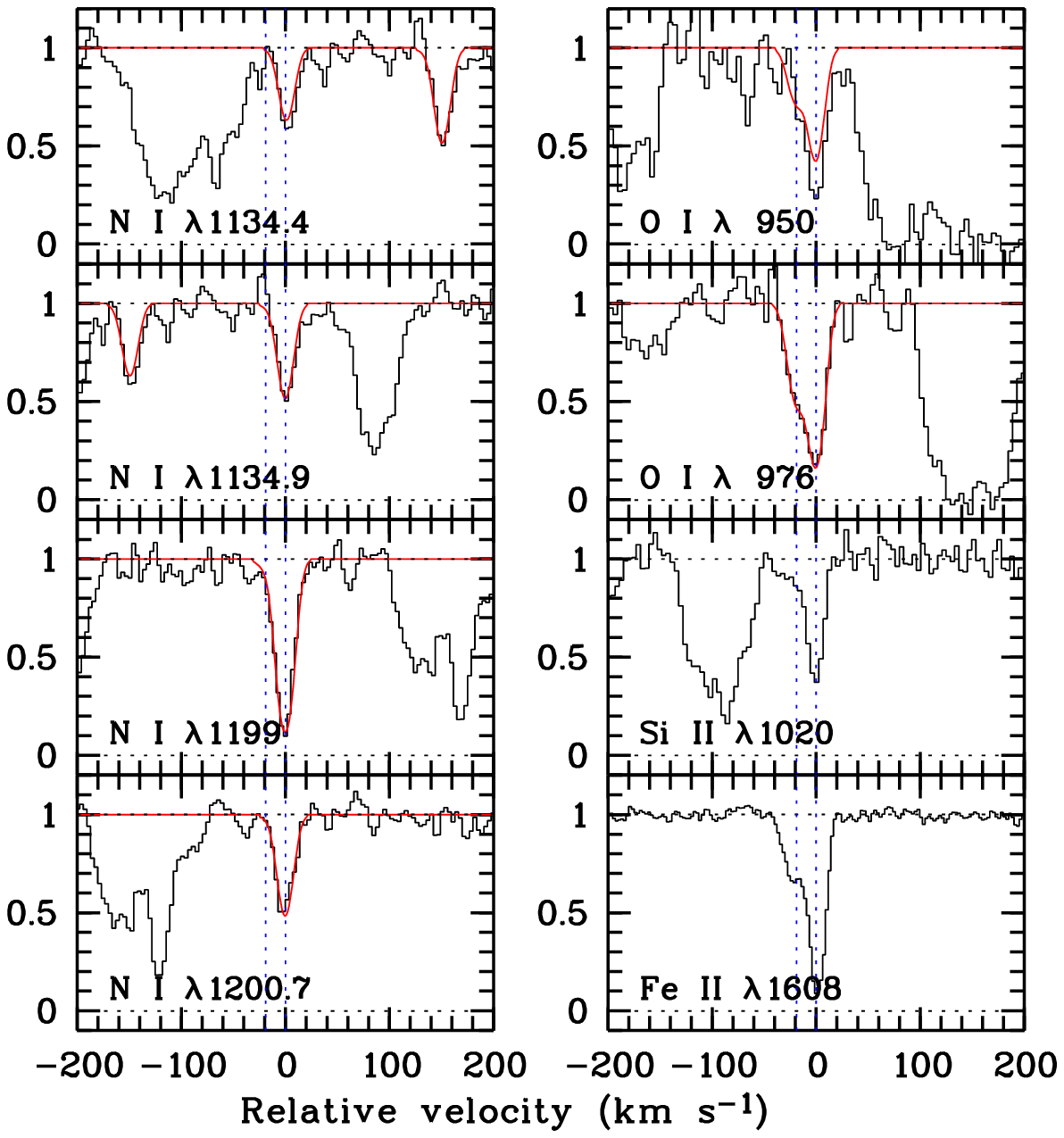}
   \includegraphics[width=7.5cm,angle=0,clip]{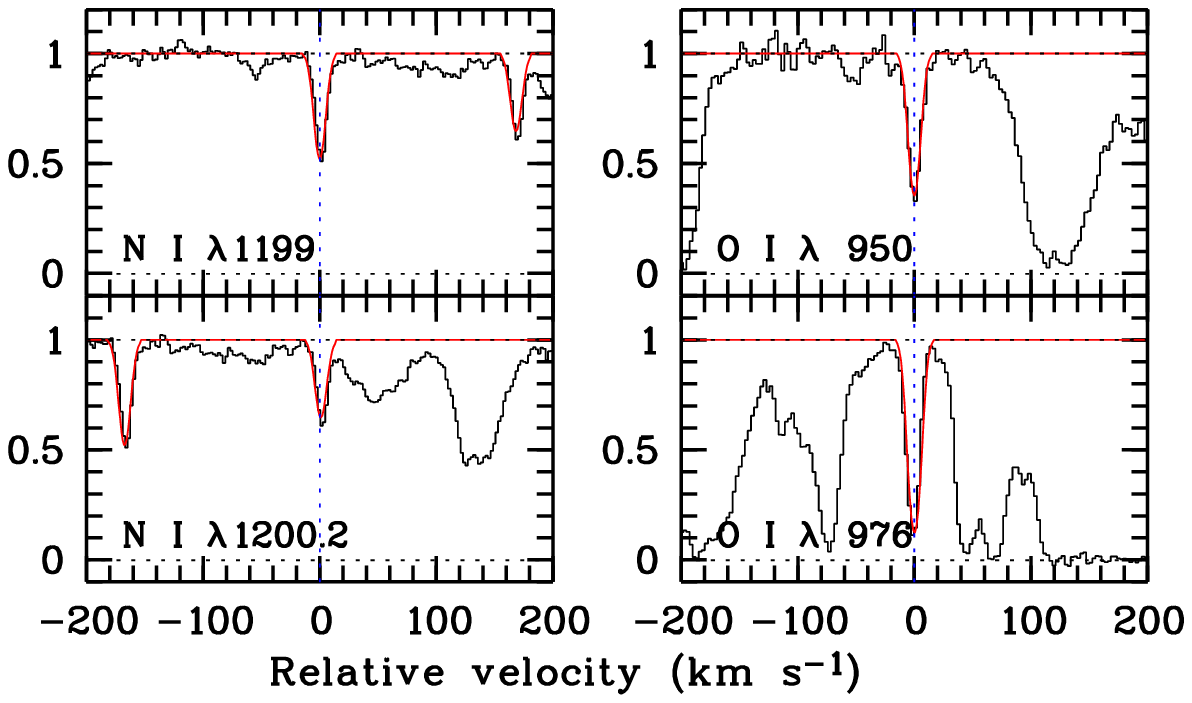}
}
   \caption{N~{\sc i} and O~{\sc i} absorption lines in the DLA systems:
Q~1108$-$077, $z$~=~3.608;
Q~1337$+$113, $z$~=~2.508; Q~1337$+$113, $z$~=~2.796 and 
Q~1340$-$136, $z$~=~3.118 
from top to bottom respectively. 
Model fits are overplotted.
\label{Spectres2}} 
    \end{figure}
\begin{figure}
   \centering
\vbox{ 
\includegraphics[width=7.5cm,angle=0,clip]{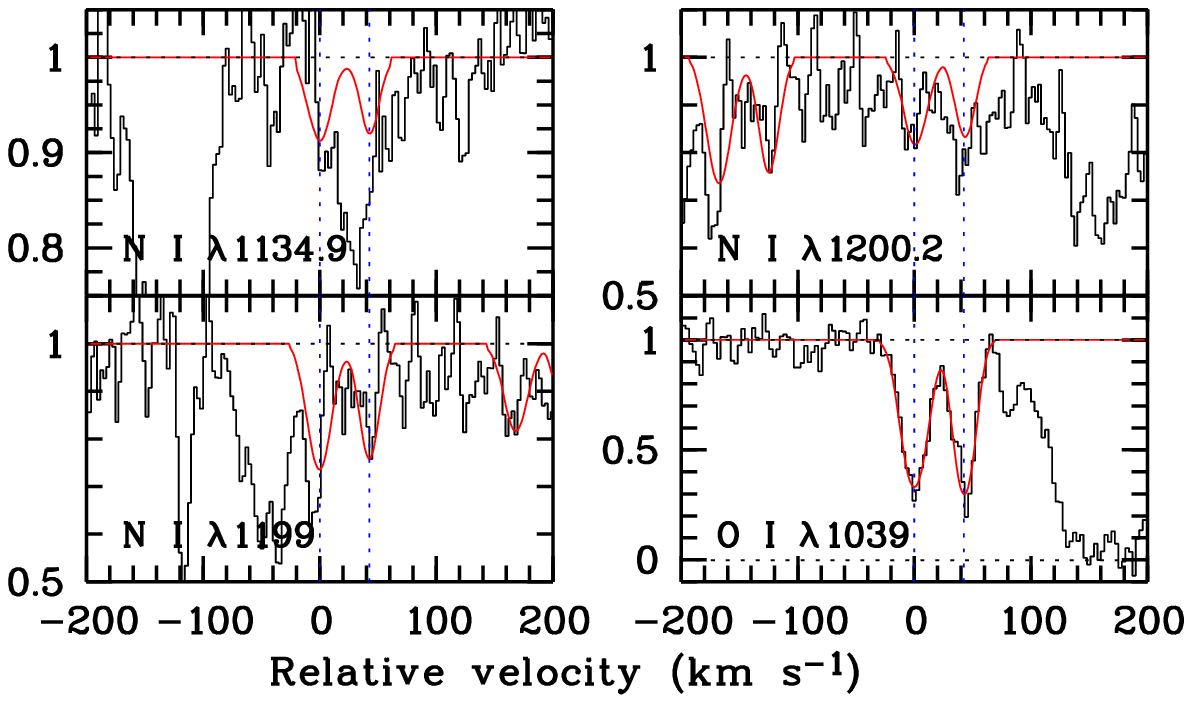}
   \includegraphics[width=7.5cm,angle=0,clip]{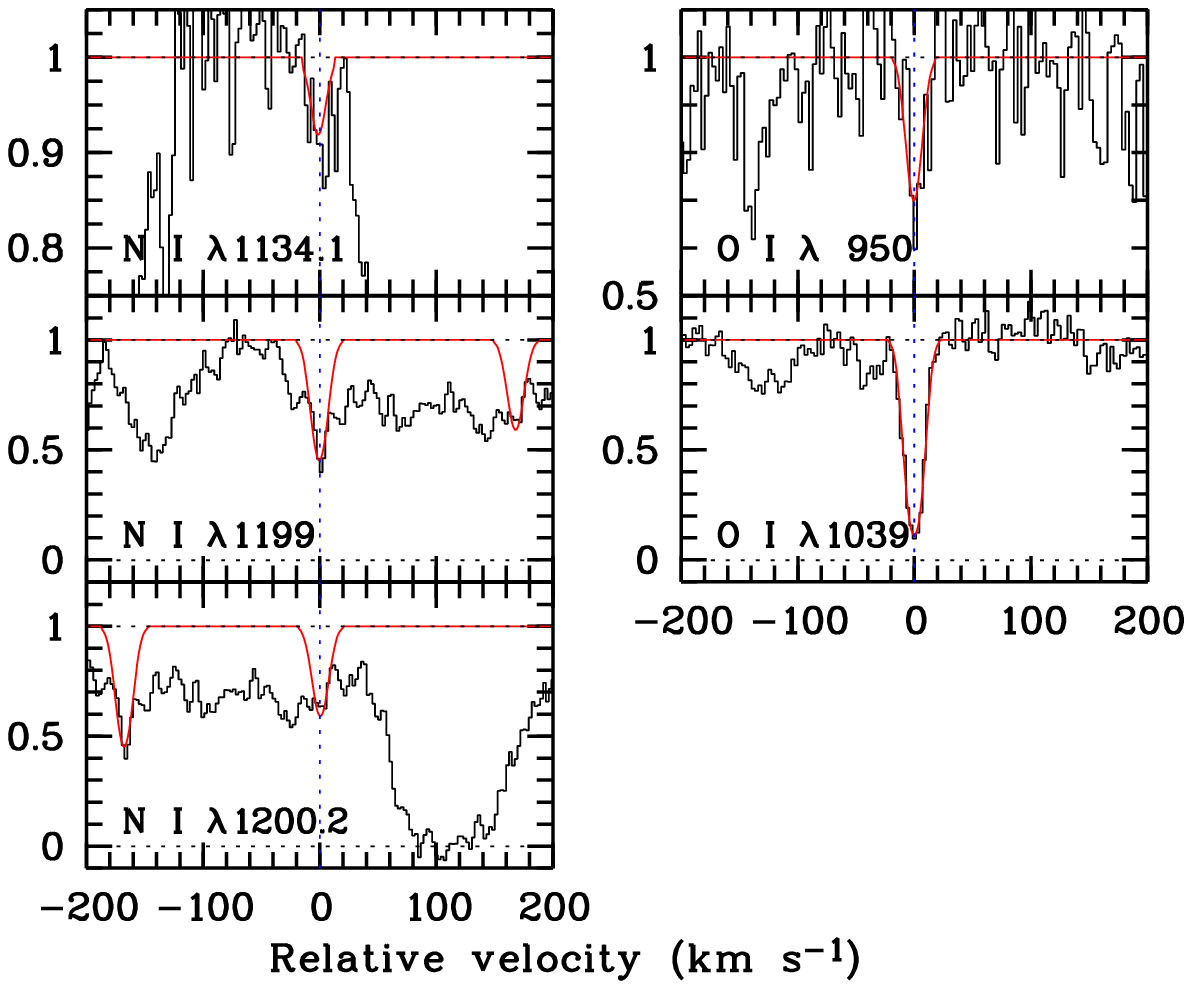}
   \includegraphics[width=7.5cm,angle=0,clip]{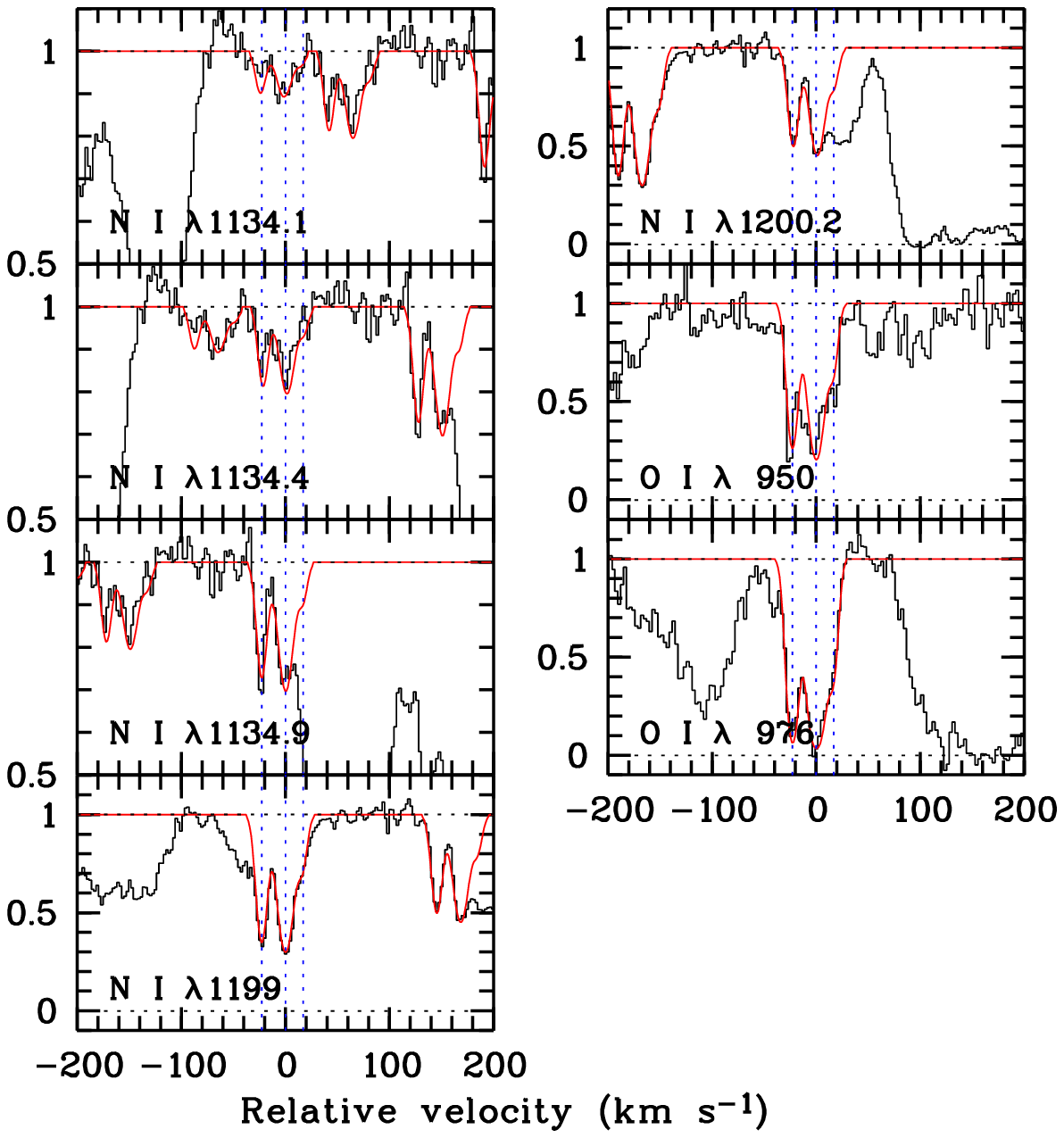}
   \includegraphics[width=7.5cm,angle=0,clip]{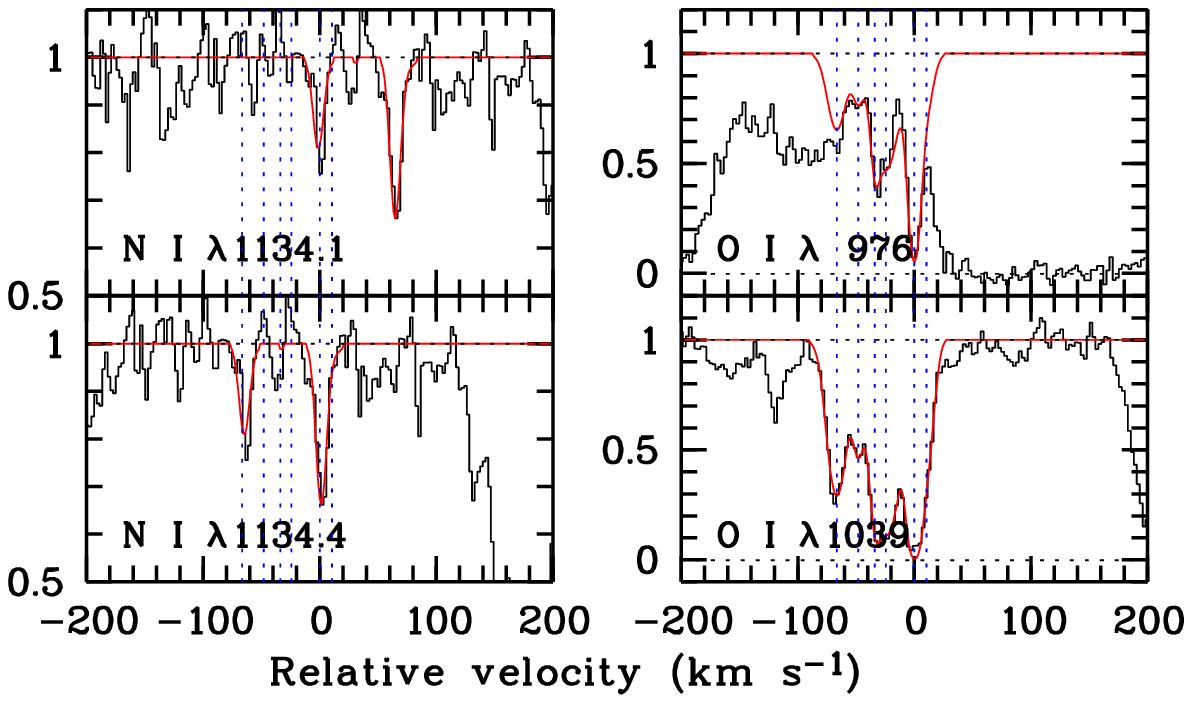}
}
   \caption{N~{\sc i} and O~{\sc i} absorption lines in the DLA systems:
Q~1409$+$095, $z$~=~2.456; Q~1409$+$095, $z$~=~2.668; Q~2059$-$360, $z$~=~3.083 and
Q~2332$-$094, $z$~=~3.057 from top to bottom respectively. 
Model fits are overplotted.
\label{Spectres3}} 
    \end{figure}

\section{Conclusion}
We have measured the [O/Fe] and [N/O] abundance ratios in 13 DLA
systems in the range $-$2.7~$<$~[Fe/H]~$<$~$-1$ and added three systems
from the literature.
\par\noindent 
%This is the first systematic survey for Oxygen abundance in DLAs.
We show that the scatter in the [O/Fe] measurements is small around
a mean value of [O/Fe]~$\sim$~0.32$\pm$0.10. This small scatter is probably 
evidence for similar nucleosynthetic histories and efficient mixing of the gas. 
In the same range of metallicity,
[O/Fe] in metal-poor stars is found to be about 0.7~dex if the measurements
are not corrected for 3D effects and 0.4~dex if corrected (Cayrel et al. 2004). 
%Given uncertainties involved, the two results should not be considered
%as interestingly similar. 
\par\noindent
In the [N/O] versus [O/H] diagram, we find that about a third of the measurements 
are located close to but below the so-called local primary plateau. 
If a plateau is to be defined for DLAs, its position is at
[N/O]~$\sim$~$-$0.9 that is lower than the plateau measured locally
in metal-poor dwarf galaxies, [N/O]~$\sim$~$-$0.57 to $-$0.74. 
%However the difference
%may be explained by the slight over-abundance of Oxygen compared to 
%Iron that we observe ([O/Fe]~$\sim$~0.32). 
All our measurements are located below the local plateau which is not the case for measurements in
metal-poor halo stars (Spite et al. 2005):
about half of the stars have [N/O]~$>$~$-$0.9. However this may be due to small 
number statistics as a quarter of the DLA systems compiled by Henry \& Prochaska (2007) have
a [N/Si] ratio larger than the value corresponding to the primary plateau 
for these elements.
We find difficult to confirm the presence of a second plateau at lower value as claimed by
Centuri\'on et al. (2003) although the [N/O] distribution is probably
double peaked around [N/O]~$\sim$~$-$0.9 and [N/O]~$\sim$~$-$1.35.
This bimodality in the [N/O] distribution could be due to small number statistics as well
% in particular because we find, contrary to previous
%studies, that the number of systems around these values are about equal. 
although a possible bimodality in the [N/$\alpha$] distribution has already been noticed 
by Prochaska et al. (2002) 
and Centuri\`on et al. (2003); see also Henry \& Prochaska (2007). 
\par\noindent
Oxygen is mainly produced in short-lived massive stars and released into
the ISM by Type II Supernovae explosions. Nitrogen is mainly produced
in long-lived intermediate mass stars and ejected in the ISM by stellar winds. 
This implies delays in the ejection of Nitrogen into the ISM even if it is
primary. This is probably why most of the DLA measurements are below the local
primary plateau. We note that the approximately constant observed [O/Fe] ratio does not
support the suggestion by Pettini et al. (2002) that Iron should
have the same evolutionary time-scale as Nitrogen so that
DLAs deficient in Nitrogen should also be deficient in Iron.
However, their argument holds only for part of the Iron production,
the part produced by Type Ia Supernovae that have the same time-scale as 
intermediate-mass stars. Indeed, models with constant star-formation over a large range of
duration produce a [Si/Fe]~$\sim$~0.3 ratio about constant over a large
metallicity range (Henry \& Prochaska 2007). These models cannot reproduce the scatter
in the measurements but one could easily claim that constant star-formation rate is
a simplistic assumption and that scatter can arise from different
star-formation histories (see also Moll\'a et al. 2006).
\par\noindent
As a consequence, considering an isolated galaxy, 
%the ISM Nitrogen abundance 
%increases (and so does the [N/O] ratio) during long periods of quiescence. 
the [N/O] ratio could decrease sharply during a starburst when Oxygen is released by SNe
(e.g. Contini et al. 2002). The scatter in the [N/O] ratio measurements could then be 
due to the intensity of the different bursts of star formation, explaining the 
different [N/O] values. After some delay, corresponding to
the lifetime of intermediate mass stars, the [N/O] ratio could increase
because of the release of most of the Nitrogen. 
%In that case, Nitrogen should
%be a primary element at these metallicities. 
The bimodality of the [N/O] 
distribution and the larger number of systems with low [N/O] ratio
are a consequence of the delay between the releases of Oxygen and 
Nitrogen, because the duration of the release is small compared to
the life-time of the stars (see also Ledoux et al. 2006b). 
It is clear that increasing the sample size would be important to improve 
our knowledge on these issues.

\begin{acknowledgements}
We thank the referee, Paolo Molaro, for a thorough reading of the
manuscript and useful comments. RS and PPJ gratefully acknowledge 
the Indo-French Centre for the Promotion of Advanced Research
(Centre Franco-Indien pour la Promotion de la Recherche Avanc\'ee)
under contract No. 3004-3. We thank Elisabeth Flam for useful discussions.
%PPJ thanks ESO-Vitacura for hospitality during the time
%part of this work has been completed.

\end{acknowledgements}

%----------------------------------------------------------------

\end{document}